\documentclass{emulateapj}

\shorttitle{AGN X-ray Emission and Black Holes}
\shortauthors{Kelly et al.}

\begin{document}

  \title{Observational Constraints on the Dependence of Radio-Quiet
    Quasar X-ray Emission on Black Hole Mass and Accretion Rate}

\author{Brandon C. Kelly, Jill Bechtold, Jonathan R. Trump, Marianne Vestergaard\altaffilmark{1}}
\affil{Steward Observatory, University of Arizona, 933 N Cherry Ave., 
  Tucson, AZ 85710}
\email{bkelly@as.arizona.edu}
\and
\author{Aneta Siemiginowska}
\affil{Harvard-Smithsonian Center for Astrophysics, 60 Garden Street, 
  Cambridge, MA 02138}
\altaffiltext{1}{Dept. of Physics and Astronomy, Robinson Hall, Tufts University, 
  Medford, MA 02155 (present address)}

\begin{abstract}
In this work we use a sample of 318 radio-quiet quasars (RQQ) to
investigate the dependence of the ratio of optical/UV flux to X-ray
flux, $\alpha_{\rm ox}$, and the X-ray photon index, $\Gamma_X$, on
black hole mass, UV luminosity relative to Eddington, and X-ray
luminosity relative to Eddington. Our sample is drawn from the
literature, with X-ray data from \emph{ROSAT} and \emph{Chandra}, and
optical data mostly from the SDSS; 153 of these sources have estimates
of $\Gamma_X$ from \emph{Chandra}. We estimate $M_{BH}$ using standard
estimates derived from the H$\beta$, Mg II, and C IV broad emission
lines. Our sample spans a broad range in black hole mass ($10^6
\lesssim M_{BH} / M_{\odot} \lesssim 10^{10}$), redshift ($0 < z <
4.8$), and luminosity ($10^{43} \lesssim \lambda L_{\lambda}
(2500$\AA$) [{\rm erg\ s^{-1}}] \lesssim 10^{48}$). We find that
$\alpha_{\rm ox}$ increases with increasing $M_{BH}$ and $L_{UV} /
L_{Edd}$, and decreases with increasing $L_X / L_{Edd}$. In addition,
we confirm the correlation seen in previous studies between $\Gamma_X$
and $M_{BH}$ and both $L_{UV} / L_{Edd}$ and $L_X / L_{Edd}$; however,
we also find evidence that the dependence of $\Gamma_X$ of these
quantities is not monotonic, changing sign at $M_{BH} \sim 3 \times
10^8 M_{\odot}$. We argue that the $\alpha_{\rm ox}$ correlations
imply that the fraction of bolometric luminosity emitted by the
accretion disk, as compared to the corona, increases with increasing
accretion rate relative to Eddington, $\dot{m}$. In addition, we argue
that the $\Gamma_X$ trends are caused by a dependence of X-ray
spectral index on $\dot{m}$. We discuss our results within the context
of accretion models with comptonizing corona, and discuss the
implications of the $\alpha_{\rm ox}$ correlations for quasar
feedback. To date, this is the largest study of the dependence of RQQ
X-ray parameters on black hole mass and related quantities, and the
first to attempt to correct for the large statistical uncertainty in
the broad line mass estimates.
\end{abstract}


\section{INTRODUCTION}

\label{s-intro}

The extraordinary activity associated with quasars involves accretion
onto a supermassive black hole (SMBH), with the UV/optical emission
arising from a geometrically thin, optically thick cold accretion disk
\citep{shakura73}, and the X-ray continuum arising from a hot,
optically thin corona that Compton upscatters the disk UV photons
\citep[e.g.,][]{haardt91}. In highly accreting objects, like quasars
\citep[$0.01 \lesssim L_{bol} / L_{Edd} \lesssim 1$,
e.g.,][]{woo02,vest04,mclure04,koll06}, the X-ray plasma geometry is
expected to be that of a hot, possibly patchy, ionized `skin' that
sandwiches the cold disk
\citep[e.g.,][]{bis77,liang77,nayak00}. However, the evidence for this
is not conclusive, and relies on data from X-ray binaries and low-$z$
sources \citep[e.g., see the dicussion by ][]{czerny03}. Other
geometries are possible, including an accretion disk that evaporates
into a hot inner flow \citep[e.g.,][]{shap76,zdz99}, or a combination
of a hot inner flow and a corona that sandwiches the disk
\citep[e.g.,][]{pout97,sob04a}. Furthermore, radiation pressure can
drive an outflow from the disk into the corona if the two are
cospatial, thus altering the physics of the corona
\citep{proga05}. Investigations of how quasar X-ray parameters depend
on black hole mass, $M_{BH}$, and accretion rate relative to
Eddington, $\dot{m}$, offer important constraints on models of the
disk/corona system.

There have been attempts to link the evolution of SMBHs to analytic
and semi-analytic models of structure formation
\citep[e.g.,][]{kauff00,hatz03,bromley04}, where black holes grow by
accreting gas funneled towards the center during a galaxy merger until
feedback energy from the SMBH expels gas and shuts off the accretion
process \citep[e.g.,][]{silk98,fabian99,wyithe03,begel05}. This
`self-regulated' growth of black holes has recently been successfully
applied in smoothed particle hydrodynamics simulations
\citep{dimatt05,spring05}. Within this framework, the AGN or quasar
phase occurs during the episode of significant accretion that follows
the galaxy merger, persisting until feedback from the black hole
`blows' the gas away \citep[e.g.,][]{hopkins06}. Hydrodynamic
calculations have shown that line pressure is more efficient than
thermal pressure at driving an outflow \citep{proga07}, and therefore,
the efficiency of AGN feedback depends on the fraction of energy
emitted through the UV/disk component as compared to the X-ray/corona
component. If the fraction of energy emitted in the UV as compared to
the X-ray depends on $M_{BH}$ or $\dot{m}$, then it follows that the
efficiency of AGN feedback will also depend on $M_{BH}$ and
$\dot{m}$. This has important consequences for models of SMBH growth,
as the SMBH may become more or less efficient at driving an outflow
depending on its mass and accretion rate. Studies of the dependence of
quasar X-ray/UV emission on black hole mass and accretion rate are
therefore important as they allow us to constrain a $M_{BH}$- or
$\dot{m}$-dependent feedback efficiency.

Numerous previous studies have searched for a luminosity and redshift
dependence of $\alpha_{ox} = -0.384 \log L_X / L_{UV}$, the ratio of
X-ray to UV/optical flux
\citep[e.g.,][]{avni82,wilkes94,yuan98,vig03b,strat05,steffen06,kelly07c},
and $\Gamma_X$, the X-ray spectra slope
\citep[e.g.,][]{reeves00,bech03,dai04,ris05,grupe06}. Most studies
have found a correlation between $\alpha_{\rm ox}$ and UV luminosity,
$L_{UV}$, while the existence of a correlation between $\alpha_{\rm
ox}$ and $z$ is still a matter of debate
\citep[e.g.,][]{bech03,vig03b,steffen06,just07,kelly07c}. In addition,
studies of $\Gamma_X$ have produced mixed results. Some authors have
claimed a correlation between $\Gamma_X$ and luminosity
\citep[e.g.,][]{bech03,dai04} or redshift \citep[e.g.,][]{reeves97,
vig99, page03}, while, others find no evidence for a correlation
between $\Gamma_X$ and $L_{UV}$ or $z$
\citep[e.g.,][]{vig05,ris05,kelly07c}.

A correlation between $\Gamma_X$ and the $FWHM$ of the H$\beta$ line
has also been found \citep[e.g.,][]{boller96,brandt97}, suggesting a
correlation between $\Gamma_X$ and black hole mass or Eddington ratio
\citep[e.g.,][]{laor97,brandt98}. Recently, it has become possible to
obtain estimates of $M_{BH}$ for broad line AGN by calibrating results
from reverberation mapping \citep{peter04,kaspi05} for use on
single-epoch spectra
\citep{wand99,vest02,mclure02,vest06,kelly07b}. This has enabled some
authors to confirm a correlation between $\Gamma_X$ and either
$M_{BH}$ or $L_{bol} / L_{Edd}$
\citep[e.g.,][]{lu99,gier04,porquet04,picon05,shemmer06}, where the
X-ray continuum hardens with increasing $M_{BH}$ or softens with
increasing $L_{bol} / L_{Edd}$. In addition, previous work has also
found evidence for quasars becoming more X-ray quiet as $M_{BH}$ or
$L_{bol} / L_{Edd}$ increase \citep{brunner97,wang04}; however,
studies involving the dependence of $\alpha_{\rm ox}$ on $M_{BH}$ or
$L_{bol} / L_{Edd}$ have remained rare compared to studies of
$\Gamma_X$. It is important to note that the correlations inferred in
previous work generally employ broad line mass estimates in
combination with a constant bolometric correction. Therefore, most of
the correlations found in previous work are, strictly speaking,
between $\Gamma_X$ or $\alpha_{\rm ox}$ and the estimates $M_{BH}
\propto L_{\lambda}^{\gamma} FWHM^2$ and $L_{bol} / L_{Edd} \propto
L_{\lambda}^{1 - \gamma} FWHM^{-2}$, where $\gamma \sim 0.5$.

In this work, we investigate the dependence of $\alpha_{\rm ox}$ and
$\Gamma_X$ on black hole mass, optical/UV luminosity relative to
Eddington, and X-ray luminosity relative to Eddington. We combine the
main Sloan Digital Sky Survey (SDSS) sample of \citet{strat05} with
the sample of \citet{kelly07c}, creating a sample of 318 radio-quiet
quasars (RQQ) with X-ray data from \emph{ROSAT} and \emph{Chandra},
and optical spectra mostly from the SDSS; 153 of these sources have
estimates of $\Gamma_X$ from \emph{Chandra}. Because the X-ray
emission in radio-loud sources can have an additional component from
the jet \citep[e.g.,][]{zam81,wilkes87}, we focus our analysis on the
radio-quiet majority. Our sample has a detection fraction of $87\%$
and spans a broad range in black hole mass ($10^6 \lesssim M_{BH} /
M_{\odot} \lesssim 10^{10}$), redshift ($0 < z < 4.8$), and luminosity
($10^{43} \lesssim \lambda L_{\lambda} (2500$\AA$) [{\rm erg\ s^{-1}}]
\lesssim 10^{48}$), enabling us to effectively look for trends
regarding $\alpha_{\rm ox}$ and $\Gamma_X$.

The outline of this paper is as follows. In \S~\ref{s-sample} we
describe the construction of our sample, and in \S~\ref{s-specfits} we
describe the procedure we used to fit the optical continuum and
emission lines. In \S~\ref{s-mbhredd_est} we describe how we obtain
broad line mass estimates, our bolometric correction, and argue that a
constant bolometric correction provides a poor estimate of the
bolometric luminosity. In \S~\ref{s-alfox_bh} we describe the results
from a regression analysis of $\alpha_{\rm ox}$ on $M_{BH}, L_{UV} /
L_{Edd},$ and $L_X / L_{Edd}$, and in \S~\ref{s-gamx_bh} we report
evidence for a non-monotonic dependence of $\Gamma_X$ on either
$M_{BH}, L_{UV} / L_{Edd},$ and $L_X / L_{Edd}$. In
\S~\ref{s-discussion} we discuss our results within the context of AGN
disk/corona models, and we discuss the implications for a dependence
of quasar feedback efficiency on black hole mass or accretion rate. In
\S~\ref{s-summary} we summarize our main results.

We adopt a cosmology based on the the WMAP best-fit parameters
\citep[$h=0.71, \Omega_m=0.27, \Omega_{\Lambda}=0.73$,][]{wmap}. For
ease of notation, we define $L_{UV} \equiv \nu L_{\nu} (2500$\AA$),
L_X \equiv \nu L_{\nu} (2\ {\rm keV}), l_{UV} \equiv \log \nu L_{\nu}
(2500 $\AA$),$ and $m_{BH} \equiv \log M_{BH} / M_{\odot}$.

\section{SAMPLE CONSTRUCTION}

\label{s-sample}

In this analysis we combine 169 RQQs from \citet[][hereafter
K07]{kelly07c} with 149 RQQs from the main SDSS sample of
\citet[][hereafter S05]{strat05} to create a sample of 318 RQQs. Out
of these 318 sources, 276 ($86.8\%$) are detected in the X-rays. The
$z \lesssim 4$ sources from the K07 sample were selected by
cross-correlating the SDSS DR3 quasar catalogue \citep{dr3qsos} with
the {\it Chandra} public archive as of 2005 February 22. The $z
\gtrsim 4$ sources from the K07 sample consist of targeted
\emph{Chandra} RQQs taken from the literature
\citep{bech03,vig01,vig03a}, and new observations reported by K07. The
sources taken from S05 were selected from the SDSS to be contained
within the inner $19'$ of \emph{ROSAT} PSPC pointings with exposure
times $> 11$ ksec. The X-ray data for both samples are as reported by
S05 and K07.

Both the S05 and K07 samples consist only of radio-quiet quasars. We
focus our analysis on the radio-quiet majority because the radio-loud
sources have an additional component of X-ray emission arising from
the jet \citep[e.g.,][]{zam81,wilkes87,worrall87}. In addition, both
S05 and K07 omitted BAL QSOs when possible. It is necessary to remove
the BAL QSOs because their high column density gives them the
appearance of being X-ray weak \citep[e.g.,][]{green01, gall02,
gall06}, potentially biasing our analysis. However, neither S05 nor
K07 were able to remove the high-ionization BAL quasars for $z < 1.5$,
as their identification requires observations of the C IV line. In
addition, low-ionization BALs can be identified at $0.45 < z < 2.25$
based on Mg II absorption. \citet{reich03} found the fraction of BALs
in the SDSS to be $\sim 14\%$, and therefore we expect there to be $25
\pm 5$ BALs in our sample at $z < 1.5$. This number may be higher if
one relaxes the definition of a BAL quasar \citep{trump06}.

We exclude five sources from the S05 sample due to significant
intrinsic narrow UV absorption or obvious host galaxy contamination:
Source SDSS J103747.4-001643.9 ($z=1.500$) had significant C IV
absorption, and sources SDSS J124520.7-002128.1 ($z=2.354$) and SDSS
J103709.8+000235.2 ($z=2.679$) had significant absorption in both C IV
and L$\alpha$. These three sources were omitted because the absorption
prohibits obtaining an accurate line width measurement, necessary for
broad line mass estimates, and to ensure that the X-ray emission under
study is not effected by the absorption. Sources SDSS
J230440.6-082220.8 ($z = 0.201$) and SDSS J023306.0+003856.4 ($z =
0.244$) have a significant host-galaxy component in their spectra. In
addition, we exclude source SDSS J144340.8+585653.2 ($z = 4.278$) from
the K07 sample because it has significant UV absorption. We removed
source SDSS J142414.1+421400.1 ($z=1.608$) from the K07 sample and
source SDSS J170441.4+604430.5 (PG 1704+608, $z=0.372$) from the S05
sample, as both sources are radio-loud.

We could not estimate black hole masses for sources SDSS
J083206.0+524359.3 ($z=1.573$), SDSS J144231.7+011055.3 ($z=4.507$),
and PC 0910+5625 ($z=4.035$). All three of these sources are from the
K07 sample. The region containing the Mg II emission line for SDSS
J083206.0+524359.3 was missing from the SDSS spectrum, the emission
lines are too weak for SDSS J144231.7+011055.3, and an optical
spectrum was not available for PC 0910+5625.

\section{OPTICAL/UV SPECTRAL FITS}

\label{s-specfits}

Optical spectra were obtained for most sources from the SDSS. We also
obtained spectra for some of the high redshift quasars from
\citet{and01}, \citet{peroux01}, and \citet{const02}. The values of
$L_{UV}$ and $\alpha, L_{\nu} \propto \nu^{-\alpha}$, for the K07
sources are taken from K07. We processed the optical spectra for the
S05 sources in the same manner as for the K07 sources. We do this for
consistency and because S05 did not correct for quasar iron emission.

\subsection{Continuum Fitting}

\label{s-contfit}

As described by K07, we corrected the optical spectra for Galactic
absorption using the $E(B-V)$ values taken from \citet{schlegel}, as
listed in the NASA/IPAC Extragalactic Database (NED), and the
extinction curve of \citet{ccm89}, assuming a value of $A_V / E(B-V) =
3.1$. We model the continuum as a power law of the form $f_{\nu}
\propto \nu^{-\alpha}$, and the Fe emission as a scaled and broadened
iron template extracted from I Zw I. The optical iron template was
extracted by \citet{optfe}, and the UV iron template was extracted by
\citet{uvfe}. The continuum and iron emission were fit simultaneously
using the Levenberg-Marquardt method for nonlinear
$\chi^2$-minimization. Continuum flux densities were then estimated
using the power law parameters.

We were not able to use a power-law fit to calculate $L_{UV}$ for the
$z \lesssim 0.4$ sources, as the SDSS spectral range for these sources
does not contain the rest-frame UV continuum.  Instead, we use the
luminosity of the broad component of the H$\beta$ emission line,
$L_{{\rm H}\beta}$, as a proxy for $L_{UV}$. It is preferable to use
the broad H$\beta$ emission line luminosity over, say, the optical
continuum luminosity as a proxy for $L_{UV}$ because the broad
H$\beta$ emission line is not contaminated by emission from the host
galaxy, and thus should provide an approximately unbiased estimate of
$L_{UV}$. Host-galaxy contamination is likely negligible for all $z >
0.4$ sources, as $\nu L_{\nu}^* \sim 10^{44} {\rm\ ergs\ s^{-1}}$ at
$2500$\AA\ for galaxies \citep{bud05}. We fit a power-law relationship
between $L_{UV}$ and $L_{{\rm H}\beta}$ using the 44 sources at $0.4 <
z < 0.9$ for which a measurement of both quantities if available. We
used the linear regression method of \citet{kelly07a}, which allows
for measurement errors in both variables, and find
\begin{equation}
  \frac{\lambda L_{\lambda} (2500\mbox{\AA})}{10^{44}{\rm \ ergs\ s^{-1}}} = 
  (1.556 \pm 0.282) \left( \frac{L_{{\rm H}\beta}}{10^{42}{\rm \ ergs\ s^{-1}}} 
  \right)^{0.768 \pm 0.086}.
  \label{eq-luv_lhb}
\end{equation}
The intrinsic scatter about this relationship is $\approx 0.179$ dex,
implying a potential uncertainty in $l_{UV}$ inferred from this
relationship of the same magnitude. There was no trend in the
residuals with either $z$ or $L_{{\rm H}\beta}$, implying that
Equation (\ref{eq-luv_lhb}) should give unbiased estimates of $l_{UV}$
for the $z < 0.4$ sources. Values of $L_{UV}$ were estimated using
Equation (\ref{eq-luv_lhb}) for both the K07 and S05 $z < 0.4$
sources, a total of 42 sources.

\begin{figure}
  \begin{center}
    \scalebox{0.6}{\rotatebox{90}{\plotone{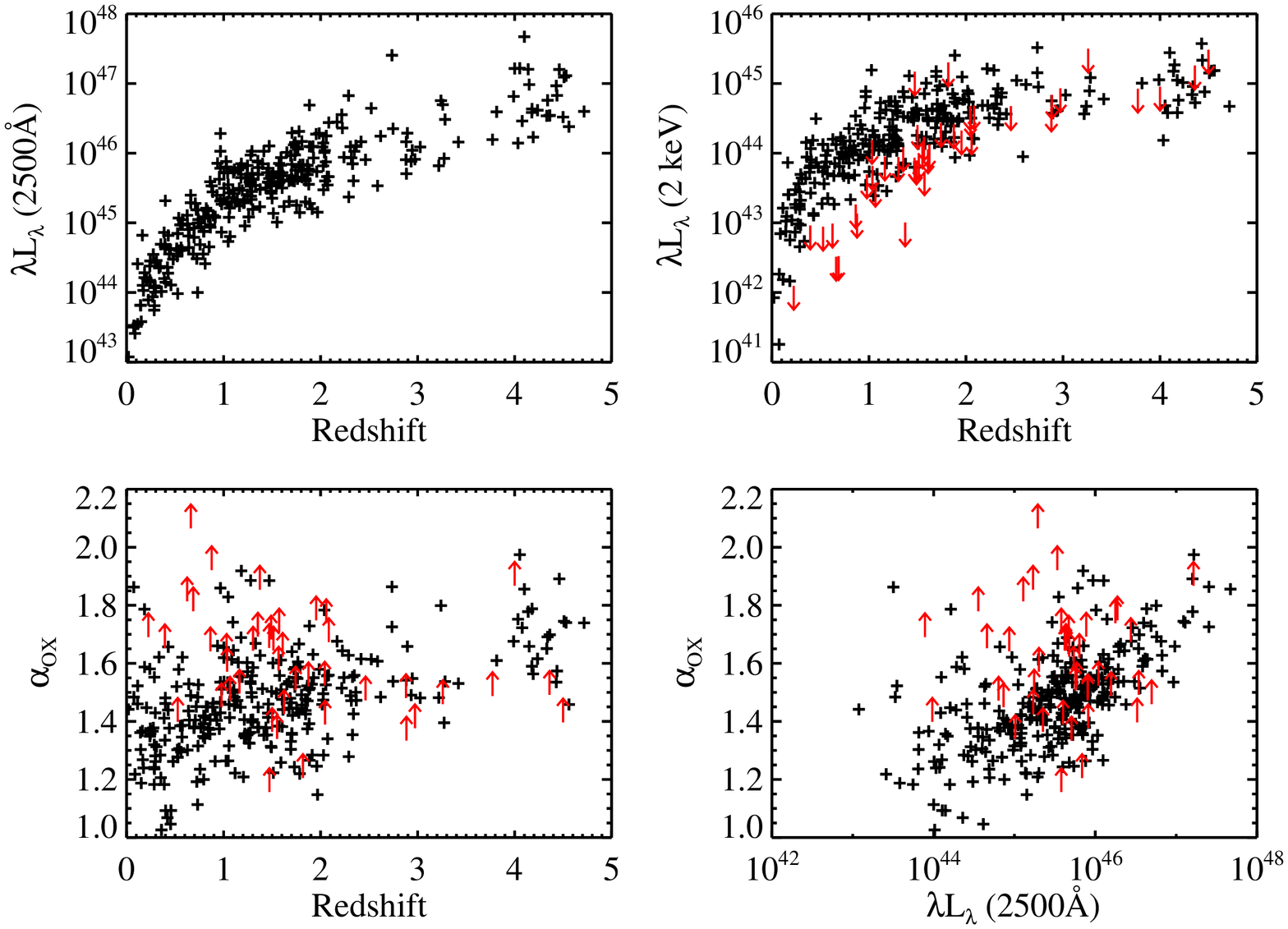}}}
    \caption{The $(L_{UV}, L_X, z)$ distribution of our
      sample. Non-detections are denoted by red
      arrows. \label{f-lum_z_dist}}
  \end{center}
\end{figure}

The distributions of $L_{UV}$ and $L_X$ as a function of
redshift are shown in Figure \ref{f-lum_z_dist}. We calculate the
ratio of optical to X-ray flux \citep{tan79} as
\begin{equation}
  \alpha_{\rm ox} = -0.384 \log (f_{\rm 2keV} / f_{2500}),
  \label{eq-alfox}
\end{equation}
where $f_{\rm 2keV}$ and $f_{2500}$ are the rest-frame flux densities at 2 keV
and $2500$\AA, respectively. If the flux density from $2500$\AA\ to 2
keV is a simple power law, then $\alpha_{\rm ox}$ is the spectral
slope of this continuum, and thus $\alpha_{\rm ox}$ may be thought of
as a crude estimate of the shape of the ionizing continuum. The
parameter $\alpha_{ox}$ is an important parameter for model
comparison, as it summarizes the amount of energy emitted in the X-ray
region (most likely a Comptonized component), compared with that
emitted in the optical-UV (accretion disk component). The distribution
of $\alpha_{\rm ox}$ as a function of $L_{UV}$ and $z$ are also
shown in Figure \ref{f-lum_z_dist}.

The error on $\alpha_{\rm ox}$ is the result of measurement errors on
the UV and X-ray flux, as well as error caused by quasar variability
over the different epochs for the UV and X-ray observations. In
general, the measurement errors on $l_{UV}$ are negligible compared to
the error on $l_X$. S05 estimates a error on the X-ray flux of $\sim
0.23$ dex, including both the contributions from measurement error and
variability. Typical long-term X-ray variability for Seyfert 1s is
$20\%$--$40\%$ with no obvious trend with luminosity \citep{grupe01,
uttley02, mark03}. The measurement errors in $l_X$ for the K07 sample
are typically $\sim 0.07$ dex. Assuming X-ray variability amplitudes
of $30\%$, this implies typical uncertainties in the X-ray luminosity
of $\sim 0.15$ dex. Therefore, we estimate the uncertainty on
$\alpha_{\rm ox}$ to be $\sim 0.06$ for the K07 sources and $\sim
0.09$ for the S05 sources.

\subsection{Line Profile Extraction and Fitting}

\label{s-lineprof}

We extracted the H$\beta$, Mg II, and C IV emission lines in order to
use their widths in our black hole mass estimates
\citep[e.g.,][]{vest02,mclure02,vest06}. These line were extracted by
first subtracting the continuum and Fe emission, interpolating over
any narrow absorption features, and modelling all lines within the
extraction region as a sum of Gaussian functions. Any nearby lines
were then subtracted, leaving only the broad emission line profile. In
all cases the line profile extraction was done interactively and every
line fit was inspected visually.

For H$\beta$, we extracted the region within $\pm 2 \times 10^4 \ {\rm
km \ s}^{-1}$ of $4861$\AA, where we use the standard convention that
negative velocities are blueward of a given wavelength. The H$\beta$
profile was modeled as a sum of 2--3 Gaussian functions. The [O III]
$\lambda 4959$\AA\ and [O III] $\lambda 5007$\AA\ lines were modeled
as a sum of 1--2 Gaussian functions, depending on the signal-to-noise
of the lines. A sum of two Gaussian functions was used for the higher
$S/N$ lines because the [O III] line profiles are not exactly a
Gaussian function; the individual Gaussian components are not
considered to be physically distinct components. The widths of the
narrow Gaussian functions for H$\beta$ and [O III] lines were fixed to
be equal to eachother. The [O III] lines and the narrow component of
the H$\beta$ line were then subtracted, leaving the broad component of
H$\beta$.
  
For Mg II, we extracted the region within $\pm 2 \times 10^4 \ {\rm km
\ s}^{-1}$ of $2800$\AA. There are no nearby non-iron emission lines
that Mg II is blended with, so the extraction is trivial after
removing the Fe and continuum emission.

For C IV, we extracted the region within $-2 \times 10^4 \ {\rm km \
s}^{-1}$ and $3 \times 10^4 \ {\rm km \ s}^{-1}$ of $1549$\AA. The C
IV line was modeled as a sum of 2--3 Gaussian functions, and He II $\lambda
1640$ and O III] $\lambda 1665$ were modeled as a sum of 1--2
Gaussian functions each. After obtaining estimates of the He II and O III]
profiles, we subtracted these components. We did not model the N IV]
$\lambda 1486$ emission line as this line is typically weak and lost
in the C IV wings.

In order to estimate $M_{BH}$, it is necessary to measure the $FWHM$
of the emission lines. After extracting the line profiles, we estimate
the $FWHM$ for the H$\beta$, Mg II, and C IV emission lines by fitting
them to a sum of 1--5 Gaussian functions, enabling us to obtain a
smooth representation of each line. In contrast to our profile
extraction technique, we choose the number of Gaussian functions to
minimize the Bayesian Information Criterion \citep[$BIC$,][]{bic}. The
$BIC$ is a common criterion to use for selecting the number of
parameters in a model \citep[e.g., see][]{hastie01}; the model that
minimizes the $BIC$ is approximately the model that is most supported
by the data. For Gaussian errors, as assumed in this work, the BIC is
simply a modification to the standard $\chi^2$ statistic:
\begin{equation}
  BIC = \chi^2 + 3K \ln n,
  \label{eq-bic}
\end{equation}
where $K$ is the number of Gaussian functions used, $3K$ is the number
of free parameters, and $n$ is the number of data points used in the
fit. Using the $BIC$ to `fit' the number of Gaussian functions thus allows us
more flexibility in obtaining a smooth representation of the line
profile, as we are not choosing the number of Gaussian functions
arbitrarily. Once a smooth representation is obtained, we
automatically measure the $FWHM$ directly from the best fit line
profile.

The standard errors on $FWHM$ are estimated using a bootstrap
method. We simulated 100 `observed' emission lines by adding random
Gaussian noise to the best fit line profile with standard deviation
equal to the noise level of the spectrum, including the propagated
errors from the continuum and iron emission fitting. We then fit each
of the simulated emission lines, keeping the number of Gaussian
functions fixed at the number found from fitting the original profile,
and measured the $FWHM$ for each simulated line. The standard error on
$FWHM$ was then estimated as the standard deviation of the $FWHM$
values measured from the simulated line profiles.

\section{ESTIMATING $M_{BH}$}

\label{s-mbhredd_est}

Recently, reverberation mapping studies of broad line AGN
\citep[e.g.,][]{peter04} established a correlation between the broad
line region (BLR) size, $R$, and the continuum luminosity \citep[the
$R$--$L$ relationship, e.g.,][]{kaspi05,bentz06}. This has made it
possible to estimate black hole virial mass $M_{BH} = f v^2 R / G$ for
individual sources, where the BLR velocity $v$ is estimated from the
width of an emission line
\citep[e.g.,][]{wand99,vest02,mclure02,vest06}. We choose the
proportionality constant to give broad line mass estimates consistent
with the $M_{BH}$--$\sigma$ relationship \citep{gebh00,merr01,trem02},
$f = 1.4 \pm 0.45$ \citep{onken04}. An estimate of the Eddington
luminosity can be computed as $L_{Edd} = 1.3 \times 10^{38} M_{BH} /
M_{\odot}\ {\rm erg\ s^{-1}}$.

\subsection{Black Hole Mass Estimates from H$\beta$, Mg II, and C IV}

\label{s-mbhest}

In this work we estimate $M_{BH}$ from the H$\beta$, Mg II, and C IV
emission lines. We use the relationship of \citet{vest06} to estimate
$M_{BH}$ from the H$\beta$ and C IV emission lines, and the datails of
the Mg II calibration will be discussed in a forthcoming paper
\citet[][in preparation]{vest08}. The calibration for the Mg II mass
estimates was calculated to ensure that they are consistent with the
mass estimates based on H$\beta$ and C IV. We have 49 sources with
both H$\beta$ and Mg II mass estimates, and 73 sources with both C IV
and Mg II mass estimates. Both samples show consistent mass estimates
between the different emission lines, within the intrinsic uncertainty
in the broad line mass estimates ($\sim 0.4$ dex).

We will denote the broad line mass estimates as $\hat{M}_{BL}$, and
$\hat{m}_{BL} \equiv \log \hat{M}_{BL} / M_{\odot}$. It is important
to distinguish between $\hat{M}_{BL}$ and $M_{BH}$, as $\hat{M}_{BL}
\propto L^{\gamma} FWHM^2, \gamma \sim 0.5,$ is an estimate of
$M_{BH}$ derived from reverberation mapping, and thus in general
$\hat{M}_{BL} \neq M_{BH}$. The statistical uncertainty needs to be
taken into account when analyzing correlations involving derived
quantities like $\hat{M}_{BL} \propto L^{\gamma} FWHM^2$, as they can
bias the results \citep{kelly07b,kelly07a}.

The uncertainty in $f$ increases the formal statistical uncertainty in
the broad line estimates of $M_{BH}$ to $\sim 0.46$ dex. Our adopted
formal uncertainty of $\sim 0.46$ dex is merely statistical, and
additional systematic uncertainties in reverberation mapping may
contribute \citep{krolik01,collin06}. Because our adopted uncertainty
of $\sim 0.46$ describes the scatter in broad line mass estimates
about the reverberation mapping estimates, as calibrated via the
$M_{BH}$--$\sigma$ relationship, the regression results found in this
work should be understood as results that could have been obtained if
we had reverberation-based $M_{BH}$ for the sources in this
work. However, instead of reverberation-based mass estimates, we have
broad line mass estimates with `measurement error' equal to $\sim
0.46$ dex with respect to the reverberation-based mass estimates, thus
increasing the uncertainty from the regression analysis.

For most sources, measurement errors on $FWHM$ and $L_{\lambda}$ did
not significantly contribute to the uncertainty on $M_{BH}$. If there
were two emission lines in the same spectrum we averaged the two mass
estimates, where the average was weighted by the uncertainties in the
two estimates. The distribution of $M_{BH}$ as a function of $z$ for
our sample is shown in Figure \ref{f-mbh_vs_z}. The broad line masss
estimates for the sources in our sample are reported in Table
\ref{t-sample}.

\subsection{Eddington Ratio Estimates}

\label{s-eddrat}

A constant bolometric correction has been used in most previous
studies involving the AGN Eddington ratio. However, recent work by
\citet{vasud07} has suggested that bolometric corrections show a large
spread with no obvious dependence on luminosity. Furthermore, these
authors found evidence that the bolometric correction depends on the
Eddington ratio. This implies that the error in the bolometric
correction is correlated with Eddington ratio, which therefore implies
that the error in the estimated Eddington ratio is correlated with the
actual Eddington ratio. An Eddington ratio-dependent error in the
bolometric correction my cause problems when using the estimated
Eddington ratios to infer correlations.

\begin{figure}
  \begin{center}
    \includegraphics[scale=0.33,angle=90]{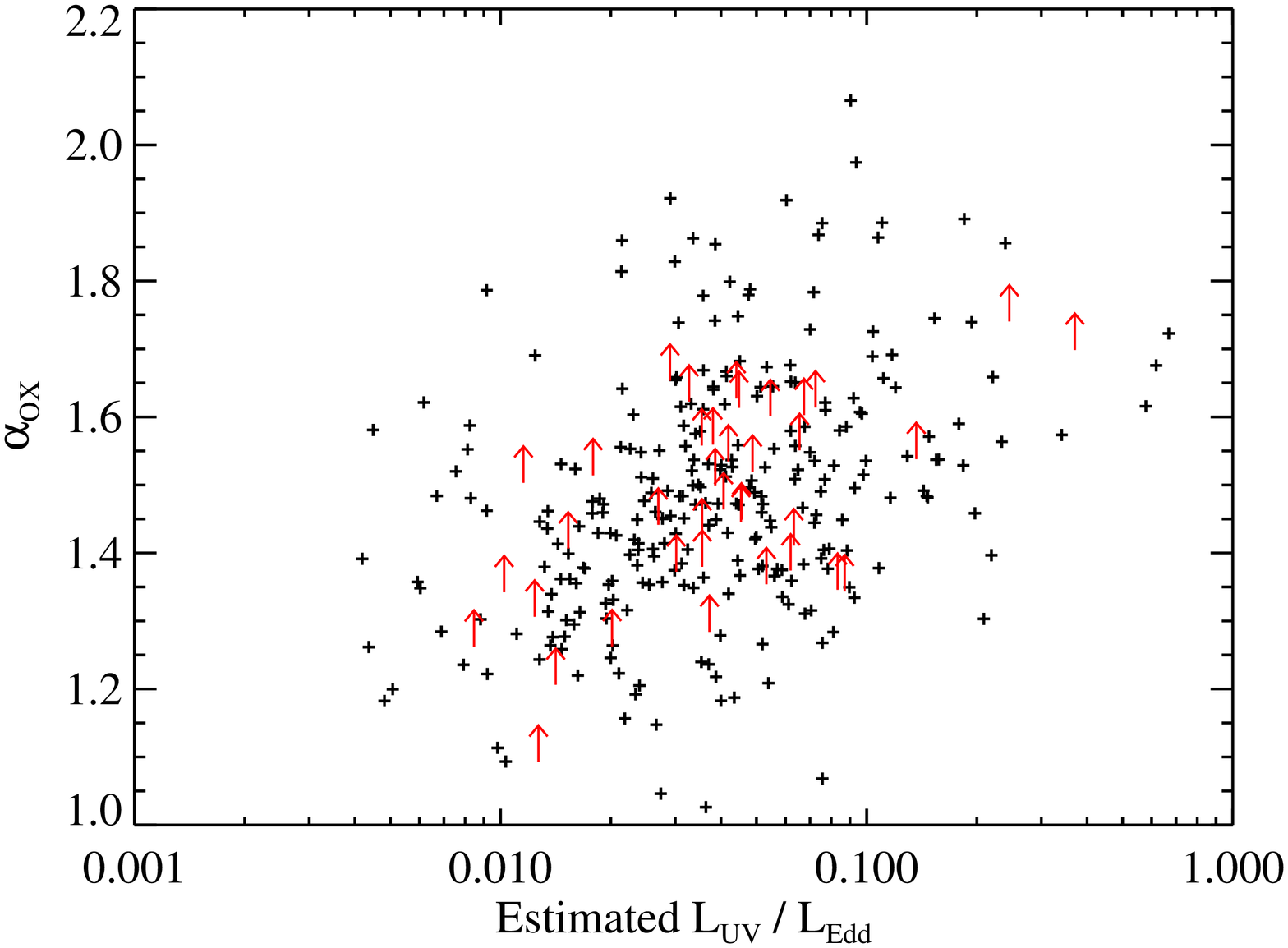}
    \includegraphics[scale=0.33,angle=90]{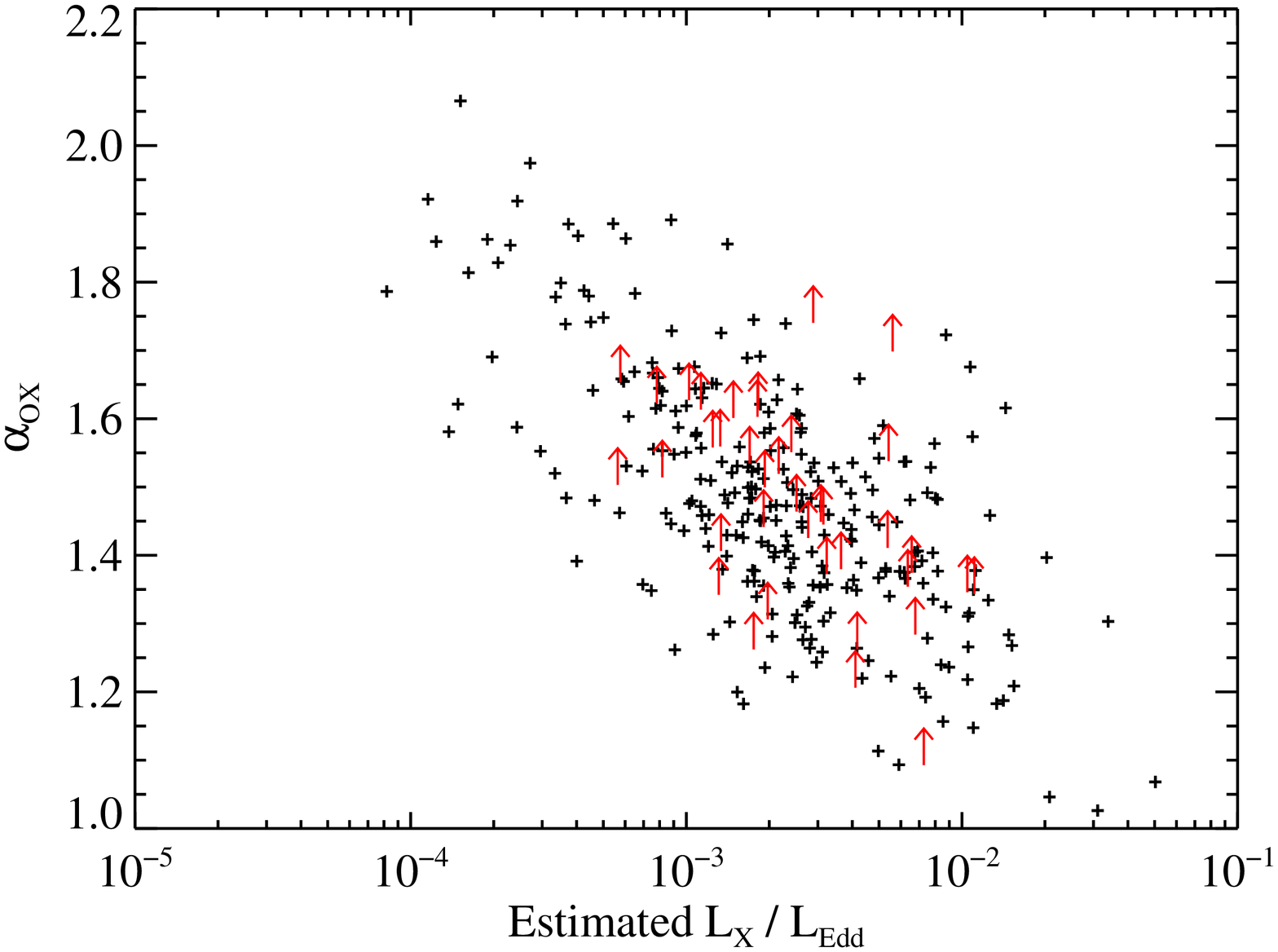}
    \caption{The distribution of $\alpha_{\rm ox}$ as a function of
    $L_{UV} / L_{Edd}$ (left) and $L_X / L_{Edd}$ (right). The
    opposite dependence of $\alpha_{\rm ox}$ on $L_{UV} / L_{Edd}$ and
    $L_X / L_{Edd}$ suggests that at least one of these quantities is
    not proportional to $L_{bol} / L_{Edd}$. As such, we do not employ
    bolometric corrections in this work, and instead compare directly
    with $L_{UV} / L_{Edd}$ and $L_X / L_{Edd}$.
    \label{f-alfox_redd}}.
  \end{center}
\end{figure}

Further difficulties with a constant bolometric correction are
illustrated with Figure \ref{f-alfox_redd}. In \ref{f-alfox_redd} we
plot $\alpha_{\rm ox}$ as a function of $L_{UV} / \hat{L}_{Edd}$ and
$L_X / \hat{L}_{Edd}$, where we estimate the Eddington luminosity from
the broad line mass estimates as $\hat{L}_{Edd} = 1.3 \times 10^{38}
\hat{M}_{BL} / M_{\odot}\ {\rm erg\ s^{-1}}$. As with $\hat{M}_{BL}$,
we use the notation $\hat{L}_{Edd}$ to emphasize that $\hat{L}_{Edd}$
is an estimate of the true $L_{Edd}$ based on the broad line mass
estimates, and therefore $\hat{L}_{Edd} \propto L^{\gamma}
FWHM^2$. Constant bolometric corrections are often applied to either
the optical/UV or X-ray luminosity. If a constant bolometric
correction was valid for both $L_{UV}$ and $L_X$, then we would expect
that $L_{UV} / \hat{L}_{Edd} \propto L_X / \hat{L}_{Edd} \propto
L_{bol} / L_{Edd}$. However, while a correlation between $\alpha_{\rm
ox}$ and both $L_{UV} / \hat{L}_{Edd}$ and $L_X / \hat{L}_{Edd}$ is
apparent, they are of opposite sign. Because the correlations are of
opposite sign, it cannot be true that both $L_{UV} / \hat{L}_{Edd}$
and $L_X / \hat{L}_{Edd}$ are proportional to the Eddington ratio,
$L_{bol} / L_{Edd}$.

Because of the current significant uncertainty regarding RQQ
bolometric corrections, we take the conservative approach and merely
compare $\alpha_{\rm ox}$ and $\Gamma_X$ with $L_{UV} / L_{Edd}$ and
$L_X / L_{Edd}$. The estimated values of $L_{UV} / L_{Edd}$ and $L_X /
L_{Edd}$ for the sources in our sample are reported in Table
\ref{t-sample}. We can write $L_{UV} / L_{Edd} \propto f_{UV} L_{bol}
/ L_{Edd}$ and $L_X / L_{Edd} \propto f_X L_{bol} / L_{Edd}$, where
$f_{UV}$ and $f_X$ are the inverses of the bolometric corrections for
$L_{UV}$ and $L_X$, respectively. The quantities $f_{UV}$ and $f_X$
are proportional to the fraction of the bolometric luminosity emitted
at 2500\AA\ and 2 keV. Then, correlations between either $\alpha_{\rm
ox}$ or $\Gamma_X$ and $L_{UV} / L_{Edd}$ will result if $\alpha_{\rm
ox}$ or $\Gamma_X$ is correlated with $f_{UV}$, the Eddington ratio,
or both, and likewise for $f_X$.

\begin{figure}
  \begin{center}
    \scalebox{0.6}{\rotatebox{90}{\plotone{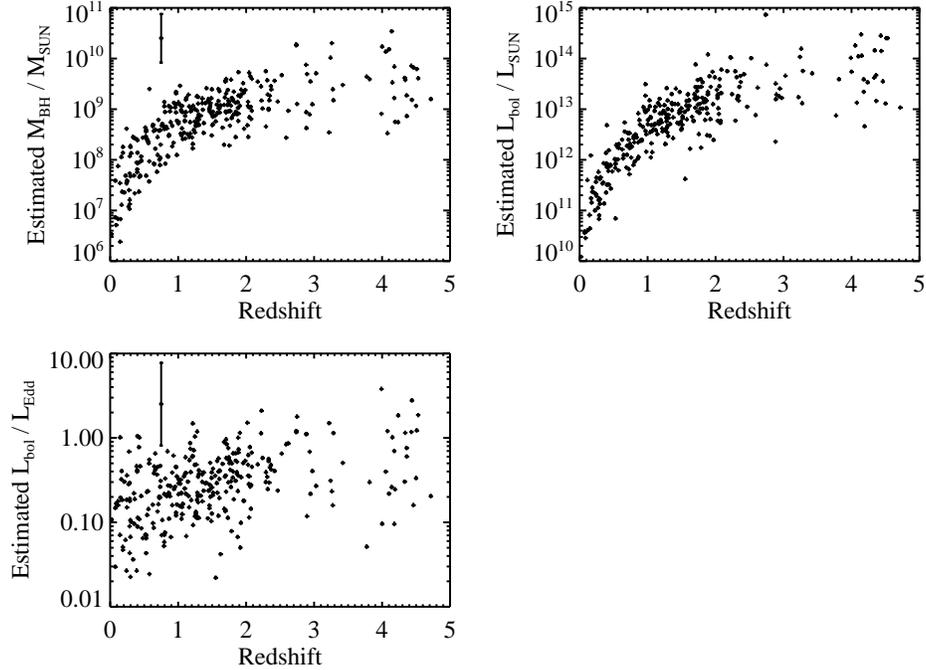}}}
    \caption{The distributions of estimated $M_{BH}, L_{bol},$ and
    $L_{bol} / L_{Edd},$ as a function of $z$ for our sample. The data
    points with error bars in the left two plots are fictitious data
    points illustrating the typical error in $\hat{M}_{BL}$ and
    $\hat{L}_{bol} / \hat{L}_{Edd}$. \label{f-mbh_vs_z}}
  \end{center}
\end{figure}

While we do not use an estimate of the Eddington ratio in our
analysis, it is helpful to estimate the distribution of Eddington
ratios probed by our sample. We assume the bolometric correction
described in \citet{hopkins07} for the $z < 1.5$ sources, and constant
bolometric described in \citet{vest04} of $L_{bol} = 4.62 \lambda
L_{\lambda}(1350$\AA$)$ at $z > 1.5$. In Figure \ref{f-mbh_vs_z} we
also show the distribution of estimated Eddington ratios as a function
of $z$. Because the distribution of estimated $L_{bol} / L_{Edd}$ is
the true distribution of $L_{bol} / L_{Edd}$ broadened by the
distribution of errors in the estimates, our sample likely probes a
smaller range in Eddington ratio than that inferred from Figure
\ref{f-mbh_vs_z}. Therefore, at most our sample probes RQQs with
Eddington ratios $0.03 \lesssim L_{bol} / L_{Edd} \lesssim 2$.

\section{DEPENDENCE OF $\alpha_{\rm ox}$ ON $M_{BH}, L_{UV} / L_{Edd},$ and $L_X / L_{Edd}$}

\label{s-alfox_bh}

We used our sample of 318 sources with estimates of $M_{BH}$ to
investigate the dependence of $\alpha_{\rm ox}$ at a given black hole
mass, $L_{UV} / L_{Edd},$ and $L_X / L_{Edd}$. We use linear
regression analysis in order to understand how $\alpha_{\rm ox}$
varies with respect to these parameters. We use the method of
\citet{kelly07a} to estimate the regression parameters. The method of
\citet{kelly07a} accounts for measurement errors, non-detections, and
intrinsic scatter. In addition, \citet{kelly07a} adopts a Bayesian
approach, computing the posterior probability distribution of the
parameters, given the observed data. Thus the uncertainties on the
regression coefficients have a straight-forward interpretion, and do
not rely on large-sample approximations. Many other methods, such as
traditional maximum-likelihood, assume that the errors in the
regression parameters follow a Gaussian distribution, which is valid
as the sample size approaches infinity. However, this assumption is
not necessarily valid for our finite sample size, especially in the
presence of censoring (i.e., presence of upper/lower limits) and
significant measurement error. The method of \citet{kelly07a} directly
estimates the probability distribution of the regression parameters,
and is therefore preferred.

\begin{figure}
  \begin{center}
    \scalebox{0.6}{\rotatebox{90}{\plotone{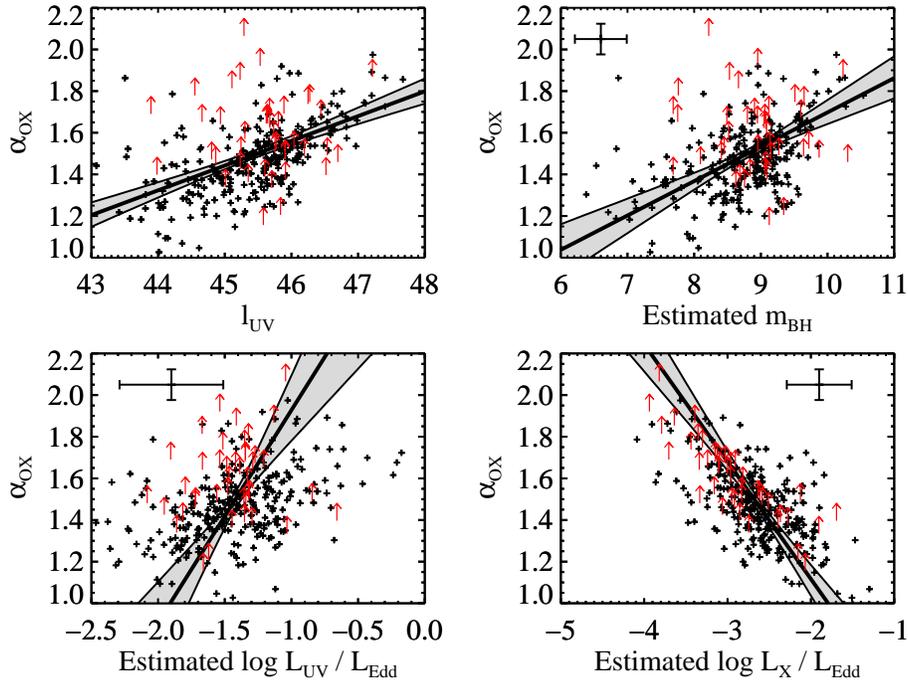}}}
    \caption{Ratio of optical/UV to X-ray flux as a function of
    $L_{UV}, M_{BH}, L_{UV} / L_{Edd},$ and $L_{X} / L_{Edd}$. The
    solid lines denote the best fit, and the shaded regions contain
    $95\%$ ($2\sigma$) of the probability on the regression line. The
    data points with error bars in the plots of $\alpha_{\rm ox}$ as a
    function of $M_{BH}, L_{UV} / L_{Edd},$ and $L_X / L_{Edd}$ are
    fictitious and illustrate the typical errors in each direction.
    \label{f-alfox_reg}}
  \end{center}
\end{figure}

We assess the simple 2-dimensional correlations between $\alpha_{\rm
ox}$ and $M_{BH}, L_{UV} / L_{Edd}$, and $L_X / L_{Edd}$, and compare
with the $\alpha_{\rm ox}$--$L_{UV}$ correlation. The results from the
regressions are
\begin{eqnarray}
  \alpha_{\rm ox} & = & -3.91^{+1.04}_{-1.01} + (0.12^{+0.02}_{-0.02}) \log L_{UV}, 
    \ \ \sigma_{\alpha_{\rm ox}} = 0.14^{+0.02}_{-0.01}, \nonumber \\
    & & \rho =  0.57^{+0.09}_{-0.10} \label{eq-alfox_luv} \\
  \alpha_{\rm ox} & = & 0.05^{+0.39}_{-0.42} + (0.17^{+0.05}_{-0.04}) \log M_{BH},
    \ \ \sigma_{\alpha_{\rm ox}} = 0.14^{+0.02}_{-0.02}, \nonumber \\
    & & \rho = 0.53^{+0.12}_{-0.13} \label{eq-alfox_mbh} \\
  \alpha_{\rm ox} & = & 2.90^{+0.71}_{-0.43} + (0.99^{+0.50}_{-0.31}) \log L_{UV} / L_{Edd},
    \ \ \sigma_{\alpha_{\rm ox}} = 0.05^{+0.05}_{-0.03}, \nonumber \\
    & & \rho = 0.95^{+0.04}_{-0.16} \label{eq-alfox_uvedd} \\
  \alpha_{\rm ox} & = & -0.03^{+0.25}_{-0.34} - (0.57^{+0.09}_{-0.12}) \log L_X / L_{Edd},
    \ \ \sigma_{\alpha_{\rm ox}} = 0.03^{+0.04}_{-0.03}, \nonumber \\
    & & \rho = -0.98^{+0.06}_{-0.02} \label{eq-alfox_xedd}
\end{eqnarray}
where $\sigma_{\alpha_{\rm ox}}$ is the intrinsic dispersion in
$\alpha_{\rm ox}$ at a given $L_{UV}, M_{BH}, L_{UV} / L_{Edd},$ or
$L_X / L_{Edd}$, $\rho$ is the linear correlation coefficient for
$\alpha_{\rm ox}$ and the respective independent variables, and the
errors are quoted at the $95\%$ $(2\sigma)$ level. All four
relationships are significant, with RQQs becoming more X-ray quiet as
$L_{UV}, M_{BH},$ or $L_{UV} / L_{Edd}$ increases, and more X-ray loud
as $L_X / L_{Edd}$ increases. Because we have attempted to correct for
the intrinsic statistical scatter in the broad line mass estimates,
Equations (\ref{eq-alfox_luv})--(\ref{eq-alfox_xedd}) refer to the
intrinsic relationships involving $M_{BH}$, barring any systematic
errors in reverberation mapping, and are not simply correlations
between $\alpha_{\rm ox}$ and the broad line mass estimates. The
estimated distributions of $\alpha_{\rm ox}$ as a function of $L_{UV},
M_{BH}, L_{UV} / L_{Edd}$, and $L_X / L_{Edd}$ are shown in Figure
\ref{f-alfox_reg}, along with the regression results.

The intrinsic dispersion in $\alpha_{\rm ox}$ quantifies the magnitude
of scatter in $\alpha_{\rm ox}$ at a given $L_{UV}, M_{BH}, L_{UV} /
L_{Edd},$ or $L_X / L_{Edd}$. Because we have attempted to account for
contribution to the scatter in $\alpha_{\rm ox}$ resulting from
measurement error and variability, $\sigma_{\alpha_{\rm ox}}$
represents the dispersion in the real physical scatter in $\alpha_{\rm
ox}$ over the population of RQQs. This `residual' scatter represents
the amount of variation in $\alpha_{\rm ox}$ that is not `explained'
by variations in $L_{UV}, M_{BH}, L_{UV} / L_{Edd},$ or $L_X /
L_{Edd}$, respectively. This intrinsic scatter in $\alpha_{\rm ox}$
may be due to variations in accretion rate, viscosity, column density,
and other quantities not included in our regression.

In \S~\ref{s-sample} we estimate that there are $25 \pm 5$ BAL quasars
in our sample at $z < 1.5$. Because these objects have the appearance
of being X-ray weak, and because redshift is artificially correlated
with luminosity and $M_{BH}$ in a flux limited sample, we expect that
the presence of unidentified BALs at $z < 1.5$ will produce an excess
of X-ray weak objects at low $L_{UV}$ and $M_{BH}$, thus flattening
the inferred slopes. Inspection of the plot of $\alpha_{\rm ox}$ and
$z$ in Figure \ref{f-lum_z_dist} suggests an excess of X-ray weak
objects at $z < 1.5$ and $\alpha_{\rm ox} \gtrsim 1.8$, implying these
objects are BAL quasars. We removed these 10 objects and refit the
regressions. Omission of these objects resulted in a steepening of the
slopes for the $L_{UV}$ and $M_{BH}$ regression, and a flattening of
the slope for the $L_{UV} / L_{Edd}$ regression. In addition, the
intrinsic dispersion in $\alpha_{\rm ox}$ decreased for the $L_{UV}$
and $M_{BH}$ regressions, while it remained the same for the $L_{UV} /
L_{Edd}$ regression. These changes were small ($\sim 10\%$) and have
no effect on our conclusions. There was no difference in the results
for the $L_X / L_{Edd}$ regression.

Once can use Equation (\ref{eq-alfox}) to express the regression
results (Eq. [\ref{eq-alfox_luv}]--[\ref{eq-alfox_xedd}]) in the
alternate form
\begin{eqnarray}
  \frac{L_{\nu}(2500\mbox{\AA})}{L_{\nu}(2\ {\rm keV})} & = &
    1.17^{+0.08}_{-0.07} \times 10^4 
    \left( \frac{\nu L_{\nu} (2500\mbox{\AA})}{10^{46} {\rm erg s}^{-1}} \right)^{0.31 \pm 0.03},
    \label{eq-lumrat_luv} \\
  \frac{L_{\nu}(2500\mbox{\AA})}{L_{\nu}(2\ {\rm keV})} & = &
    9.81^{+0.65}_{-0.63} \times 10^3 
    \left( \frac{M_{BH}}{10^{9} M_{\odot}} \right)^{0.43 \pm 0.06},
    \label{eq-lumrat_mbh} \\
  \frac{L_{\nu}(2500\mbox{\AA})}{L_{\nu}(2\ {\rm keV})} & = &
    3.51^{+15.6}_{-2.58} \times 10^7
    \left( \frac{\nu L_{\nu}(2500\mbox{\AA})}{L_{Edd}} \right)^{2.57 \pm 0.45},
    \label{eq-lumrat_uvredd} \\
  \frac{L_{\nu}(2500\mbox{\AA})}{L_{\nu}(2\ {\rm keV})} & = &
    0.85^{+1.02}_{-0.52} \left( \frac{\nu L_{\nu}(2\ {\rm keV})}{L_{Edd}}
    \right)^{-1.48 \pm 0.14},
    \label{eq-lumrat_xredd}
\end{eqnarray}
where the intrinsic dispersion in $\log L_{UV} / L_{X}$ at a given
$L_{UV}, M_{BH}, L_{UV} / L_{Edd},$ and $L_X / L_{Edd}$ is $\sim
0.356, 0.375, 0.133,$ and $0.089$ dex, respectively. In contrast to
Equations (\ref{eq-alfox_luv})--(\ref{eq-alfox_xedd}), we quote the
$68\%$ ($1\sigma$) uncertainties on the constants of proportionality,
and the posterior standard deviations on the exponents. Equations
(\ref{eq-lumrat_luv})--(\ref{eq-lumrat_xredd}) may be more physically
interpretable and allow easier comparison with models.

\section{NONMONOTONIC DEPENDENCE OF $\Gamma_X, L_{UV} / L_{Edd}$, and $L_X / L_{bol}$}

\label{s-gamx_bh}

Recent work has suggested a correlation between quasar X-ray spectral
slope, $\alpha_X = \Gamma_X - 1, f_{\nu} \propto \nu^{-\alpha_X},$ and
quasar Eddington ratio as inferred from broad line mass estimates
based on the H$\beta$ emission line
\citep[e.g.,][]{porquet04,picon05,shemmer06}. The \citet{kelly07c}
sample contains measurements of $\Gamma_X$ for 157 sources, and we
were able to estimate black hole masses for 153 of them. In this
section we use these 153 RQQs to investigate the dependence of
$\Gamma_X$ on $M_{BH}, L_{UV} / L_{Edd},$ and $L_X / L_{Edd}$.

\subsection{Regression Analysis}

\label{s-gamx_reg}

\begin{figure}
  \begin{center}
    \scalebox{0.6}{\rotatebox{90}{\plotone{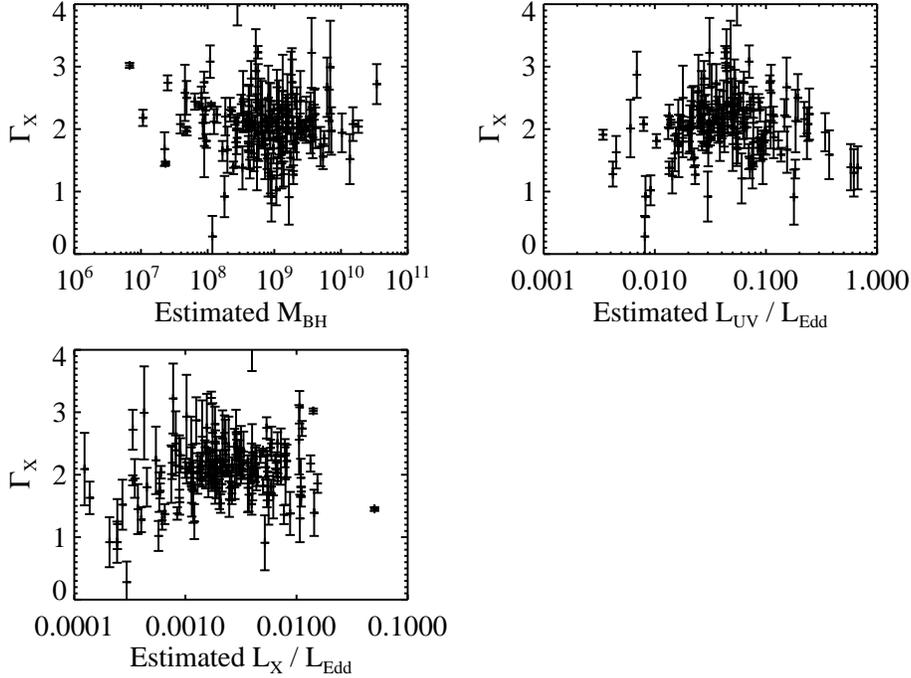}}}
    \caption{Distribution of the X-ray photon index as a function of
    estimated $M_{BH}$, $L_{UV} / L_{Edd}$, and $L_X / L_{Edd}$. For
    clarity, error bars are only shown on $\Gamma_X$, and we cut-off
    the one data point with estimated $\Gamma_X > 4$. While no obvious
    trends between $\Gamma_X$ and $M_{BH}, L_{UV} / L_{Edd},$ or $L_X
    / L_{Edd}$ exist for the whole sample, there is evidence of
    opposite trends in $\Gamma_X$ for the H$\beta$ and C IV samples.
    \label{f-gamx_vs_bh}}.
  \end{center}
\end{figure}

\begin{figure}
  \begin{center}
    \scalebox{0.6}{\rotatebox{90}{\plotone{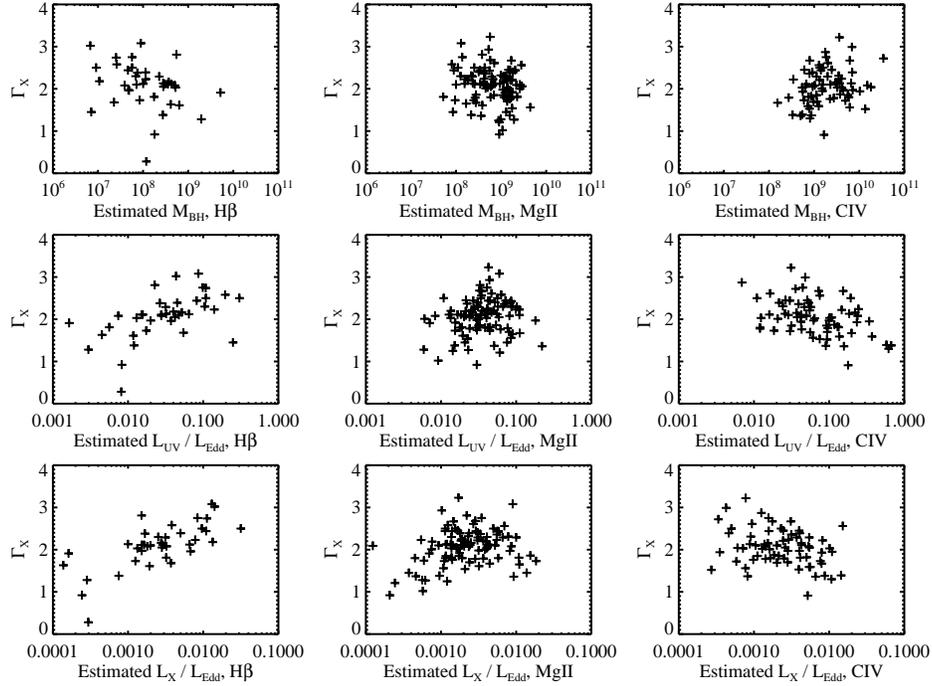}}}
    \caption{Distribution of the X-ray photon index as a function of
    estimated $M_{BH}$, $L_{UV} / L_{Edd}$, and $L_X / L_{Edd}$ for
    the individual emission lines. For clarity, error bars have been
    omitted, and we omit the one data point with estimated $\Gamma_X >
    4$. While no obvious trends exist for the whole sample, there is
    evidence of opposite trends for the H$\beta$ and C IV samples.
    \label{f-gamx_vs_bh_eline}}.
  \end{center}
\end{figure}

The distributions of $\Gamma_X$ as a function of estimated black hole
mass, $L_{UV} / L_{Edd}$, and $L_X / L_{Edd}$ are shown in Figure
\ref{f-gamx_vs_bh} for the entire sample, and in Figure
\ref{f-gamx_vs_bh_eline} seperately for each emission line. While
there does not appear to be a monotonic trend between $\Gamma_X$ and
$M_{BH}, L_{UV} / L_{Edd}$, or $L_X / L_{Edd}$ when using the entire
sample, there is evidence for a trend between $\Gamma_X$ and these
quantities when using the H$\beta$ line, and an opposite trend between
$\Gamma_X$ and these quantities when using the C IV line.

\begin{figure}
  \begin{center}
    \scalebox{0.6}{\rotatebox{90}{\plotone{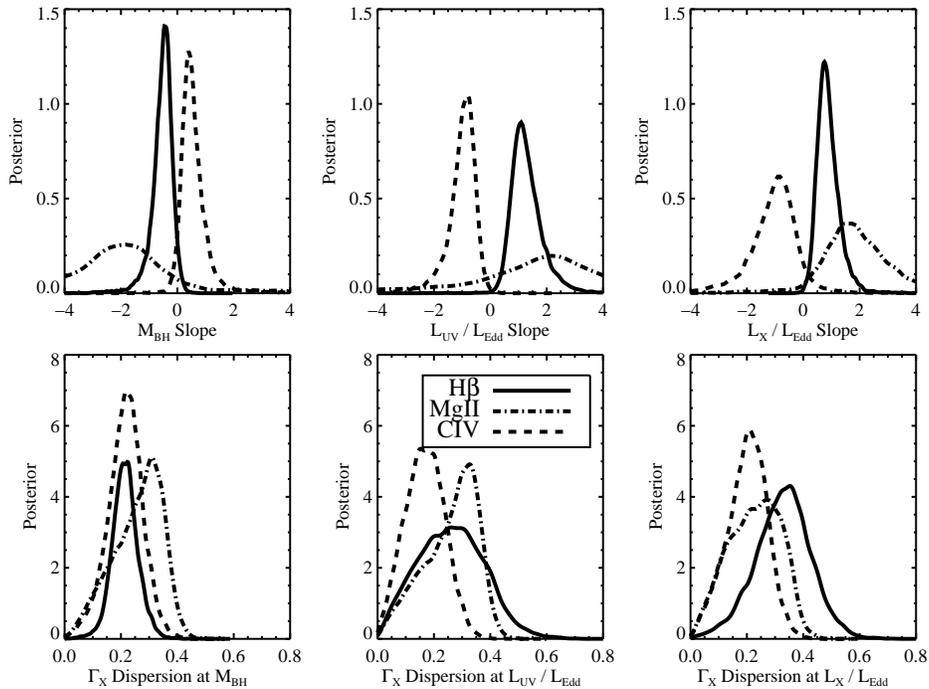}}}
    \caption{Posterior probability distributions of the slope (top)
    and intrinsic dispersion (bottom) from a linear regression of
    $\Gamma_X$ on $\log M_{BH}, \log L_{UV} / L_{Edd}$, and $\log L_X
    / L_{Edd}$. The solid lines mark the posterior for the regression
    using the H$\beta$ sample, the dashed-dotted lines mark the
    posterior for the regression using the Mg II sample, and the
    dashed lines mark the posterior for the regression using the C IV
    sample.
    \label{f-hbciv_reg}}
  \end{center}
\end{figure}

We performed a linear regression of $\Gamma_X$ on $\log M_{BH}, \log
L_{UV} / L_{Edd},$ and $\log L_X / L_{Edd}$ seperately for each
emission line. As before, we used the method of \citet{kelly07a} when
performing the regression in order to correct for the intrinsic
statistical uncertainty in the broad line estimates of $M_{BH}$. The
results for $M_{BH}$ are
\begin{eqnarray}
  \Gamma_X & = &  5.69^{+7.32}_{-4.29} - \left(0.44^{+0.53}_{-0.91}\right) \log M_{BH},\ \
    \sigma = 0.43^{+0.17}_{-0.17}, \nonumber \\
    & & \rho = -0.45^{+0.53}_{-0.44},\ {\rm (H\beta)} \label{eq-gamxmbh_hbeta} \\
  \Gamma_X & = &  -19.0^{+46.0}_{-68.0} - \left(1.92^{+7.73}_{-5.18}\right) \log M_{BH},\ \ 
    \sigma = 0.27^{+0.12}_{-0.20}, \nonumber \\
    & & \rho = -0.62^{+1.23}_{-0.36},\ {\rm (Mg II)} \label{eq-gamxmbh_mgii} \\
  \Gamma_X & = & -2.79^{+5.22}_{-10.8} + \left(0.52^{+1.17}_{-0.56}\right) \log M_{BH},\ \
    \sigma = 0.22^{+0.11}_{-0.14}, \nonumber \\
    & & \rho = 0.53^{+0.42}_{-0.58}, {\rm (CIV)} \label{eq-gamxmbh_civ}
\end{eqnarray}
the results for $L_{UV} / L_{Edd}$ are
\begin{eqnarray}
  \Gamma_X & = &  3.96^{+2.28}_{-1.15} + \left(1.23^{+1.48}_{-0.75}\right) \log L_{UV} / L_{Edd},\ \
    \sigma = 0.26^{+0.23}_{-0.21}, \nonumber \\
    & & \rho = 0.87^{+0.13}_{-0.50},\ {\rm (H\beta)} 
    \label{eq-gamxreddu_hbeta} \\
  \Gamma_X & = &  5.13^{+10.8}_{-14.4} + \left(2.14^{+7.28}_{-9.94}\right) \log L_{UV} / L_{Edd},\ \ 
    \sigma = 0.28^{+0.12}_{-0.22}, \nonumber \\
    & & \rho = 0.54^{+0.44}_{-1.40},\ {\rm (Mg II)} \label{eq-gamxreddu_mgii} \\
  \Gamma_X & = & 0.85^{+0.78}_{-2.00} - \left(0.95^{+0.65}_{-1.69}\right) \log L_{UV} / L_{Edd},\ \
    \sigma = 0.17^{+0.13}_{-0.13}, \nonumber \\
    & & \rho = -0.81^{+0.53}_{-0.18}, {\rm (CIV)} \label{eq-gamxreddu_civ}
\end{eqnarray}
and the results for $L_X / L_{Edd}$ are
\begin{eqnarray}
  \Gamma_X & = &  4.24^{+2.58}_{-1.43} + \left(0.85^{+1.01}_{-0.56}\right) \log L_X / L_{Edd},\ \
    \sigma = 0.33^{+0.19}_{-0.21}, \nonumber \\
    & & \rho = 0.76^{+0.22}_{-0.46},\ {\rm (H\beta)} \label{eq-gamxreddx_hbeta} \\
  \Gamma_X & = &  7.26^{+16.2}_{-7.13} + \left(1.97^{+5.96}_{-2.71}\right) \log L_X / L_{Edd},\ \ 
    \sigma = 0.23^{+0.15}_{-0.18}, \nonumber \\
    & & \rho = 0.77^{+0.22}_{-0.90},\ {\rm (Mg II)} \label{eq-gamxreddx_mgii} \\
  \Gamma_X & = & -0.54^{+4.05}_{-5.69} - \left(0.96^{+1.52}_{-2.17}\right) \log L_X / L_{Edd},\ \
    \sigma = 0.21^{+0.13}_{-0.15}, \nonumber \\
    & & \rho = -0.65^{+0.90}_{-0.34} {\rm (CIV)}. \label{eq-gamxreddx_civ}
\end{eqnarray}
In these equations we have quoted the errors at $95\%$ $(2\sigma)$
confidence. The probability distributions of the slope and intrinsic
dispersion are shown in Figure \ref{f-hbciv_reg}. The larger
uncertainty in the results for the Mg II sample is likely caused by
the more narrow range in $L_{UV}, L_X,$ and $M_{BH}$ probed.

\begin{figure}
  \begin{center}
    \includegraphics[scale=0.6,angle=90]{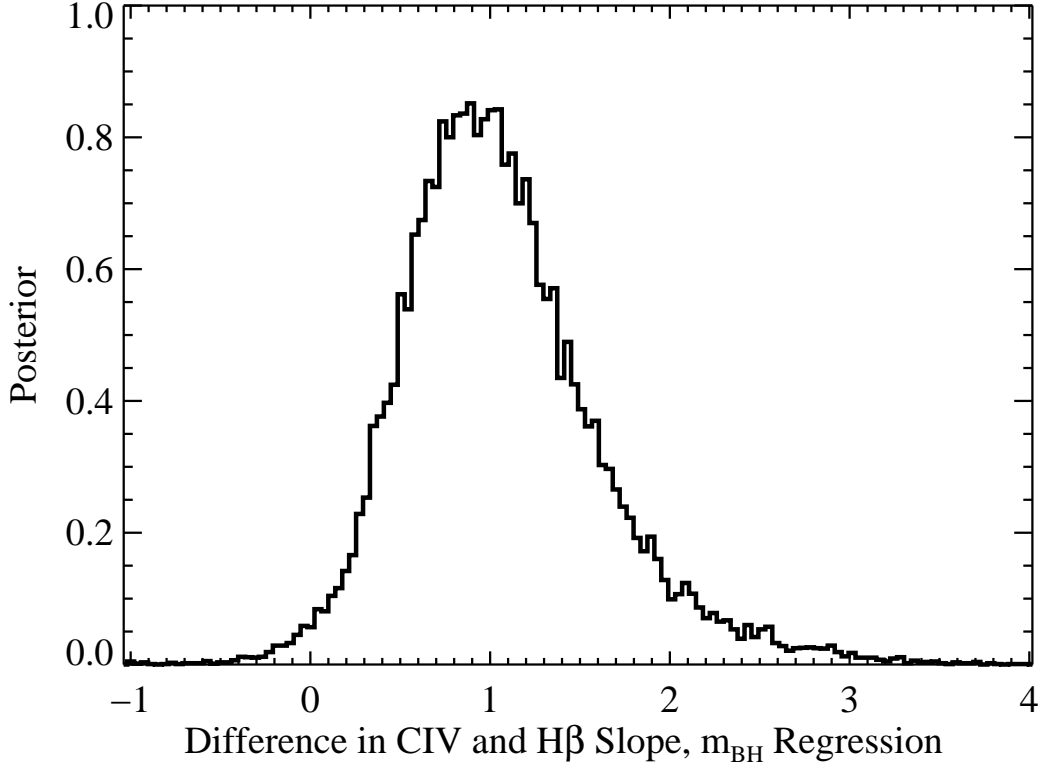}
    \caption{Posterior distribution for the difference in slopes
    between the C IV and H$\beta$ regressions of $\Gamma_X$ on
    $m_{BH}$. While there is no significant evidence that either the
    H$\beta$ or C IV regression slope is different from zero, there
    \emph{is} significant evidence that they are not the same,
    implying a nonmonotonic trend between $\Gamma_X$ and $M_{BH}$.
    \label{f-regdiff}}.
  \end{center}
\end{figure}

There are formally no significant linear correlations for the
$\Gamma_X$--$M_{BH}$ relationship. However, there is a statistically
significant difference between the H$\beta$ and C IV slopes, with
$\approx 99.3\%$ of the posterior probability at $\beta_m^{\rm CIV} >
\beta^{\rm H\beta}_m$, where $\beta_m$ denotes the $\Gamma_X$--$\log
M_{BH}$ regression slope. The probability distribution for the
difference in slopes from the $M_{BH}$ regression is shown in Figure
\ref{f-regdiff}. The significant difference in the slope for the
H$\beta$ and C IV sample implies a nonlinear relationship between
$\Gamma_X$ and $m_{BH}$, in spite of the fact that the H$\beta$ and C
IV correlations themselves are not `statistically
significant'. Results similar to the $\Gamma_X$--$\log M_{BH}$
regressions were found for the $\Gamma_X$--$L_{UV} / L_{Edd}$ and
$\Gamma_X$--$L_X / L_{UV}$ regressions, but with opposite sign and
higher statistical significance.

We performed monte carlo simulations as a consistency check on our
inferred non-monotonicity of the $\Gamma_X$ relationships. While we
have attempted to account for the significant statistical uncertainty
on the broad line mass estimates, we employ these monte carlo
simulations to ensure that the observed non-monotonic behavior is not
a spurious result caused by the uncertainty on $M_{BH}$. We performed
$10^5$ simulations under two null hypotheses: (1) that $\Gamma_X$ is
independent of $L_{UV} / L_{Edd}$, and (2) that $\Gamma_X$ depends
linearly on $\log L_{UV} / L_{Edd}$. For both cases we simulated black
hole mass estimates derived from H$\beta$ and C IV seperately. We
first simulated `true' values of $M_{BH}$ for each emission line from
a normal distribution with means equal to the observed mean of the two
respective subsamples, and variances equal to the difference between
the observed variance of the subsamples and the average intrinsic
variance in the broad line mass estimates. To simulate the uncertainty
in the mass estimates, we added random Gaussian errors to these `true'
values of $M_{BH}$ with standard deviation equal to the uncertainty in
the mass estimates, $\sim 0.4$ dex. For the case where $\Gamma_X$ was
assumed to be independent of $L_{UV} / L_{Edd}$, we simulated values
of $\Gamma_X$ from a normal distribution with mean equal to the sample
mean of $\Gamma_X$ and variance equal to the difference between the
observed variance in $\Gamma_X$ and the average of the variance in the
measurement errors. For the case where $\Gamma_X$ was assumed to
depend linearly on $\log L_{UV} / L_{Edd}$, we simulated values of
$\Gamma_X$ according to our best fit relationship to the H$\beta$
subsample, given by Equation (\ref{eq-gamxreddu_hbeta}). Finally, for
both cases we added random Gaussian errors to the simulated values of
$\Gamma_X$ by randomly reshuffling the dispersions in the measurement
errors in $\Gamma_X$.

For each of the $10^5$ simulated samples, we selected those samples
that displayed a non-monotonic trend, i.e., those sample where the
slope for the H$\beta$-based regression had a different sign from the
slope of the C IV-based regression. Under the hypothesis that
$\Gamma_X$ is independent of $L_{UV} / L_{Edd}$, only 3 of the $10^5$
simulated samples had both a non-monotonic trend and an absolute value
of the difference in slopes between the H$\beta$ and C IV regression
that were larger than that observed for our actual sample. Under the
hypothesis that $\Gamma_X$ depends linearly on $\log L_{UV} /
L_{Edd}$, none of the $10^5$ simulated samples exhibited a
non-monotonic trend. Therefore, the observed non-monotonic trend in
$\Gamma_X$ with Eddington ratio is not a spurious result caused by the
statistical uncertainty in the broad line mass estimates, in agreement
with our Bayesian regression results.

In order to investigate whether the non-monotonicity in the dependence
of $\Gamma_X$ on Eddington ratio depends on $M_{BH}$, we performed a
linear regression of $\Gamma_X$ simultaneously on $\log L_{UV} /
L_{Edd}$ and $\log M_{BH}$. This also allows us to quantify whether
the Eddington ratio is the driver behind the $\Gamma_X$--$M_{BH}$
relationship. In particular, the $\Gamma_X$--$M_{BH}$ relationship is
weak compared to the Eddington ratio relationships, and therefore it
is reasonable to conclude that Eddington ratio is the primary driver
in these relationships. We applied the multiple regression technique
of \citet{kelly07a} seperately to both the H$\beta$ and C IV
subsamples. The results are:
\begin{eqnarray}
  \Gamma_X & = &  2.58^{+4.81}_{-3.80} + 
  \left(0.18^{+0.60}_{-0.61}\right) \log M_{BH} + \nonumber \\
  & & \left(1.32^{+1.45}_{-0.80}\right) \log L_{UV} / L_{Edd},\ \
    \sigma = 0.28^{+0.22}_{-0.22}, \ {\rm (H\beta)} 
    \label{eq-gamxreddu_hbeta2} \\
  \Gamma_X & = & -4.26^{+6.45}_{-15.6} + 
  \left(0.56^{+1.44}_{-0.61}\right) \log M_{BH} - \nonumber \\
  & & \left(0.84^{+0.75}_{-2.16}\right) \log L_{UV} / L_{Edd},\ \
  \sigma = 0.17^{+0.14}_{-0.14},\ {\rm (CIV)} 
  \label{eq-gamxreddu_civ2}
\end{eqnarray}
Here, we have quoted the errors at $95\%$ significance. There is no
statistically significant evidence that $\Gamma_X$ depends on $M_{BH}$
at a given $L_{UV} / L_{Edd}$, and therefore we conclude that the
primary driver in the $\Gamma_X$ relationships is Eddington
ratio. However, this does not rule out the possibility that the
non-monotonic trends with Eddington ratio are the result of a
discontinuous change in the slopes at a `critical' $M_{BH}$, as
discussed in the next two sections.

\subsection{Is the Sign Change in the Correlations Caused by the Different 
  Emission Lines Used to Estimate $M_{BH}$?}

\label{s-eline_test}

\begin{figure}
  \begin{center}
    \scalebox{0.6}{\rotatebox{90}{\plotone{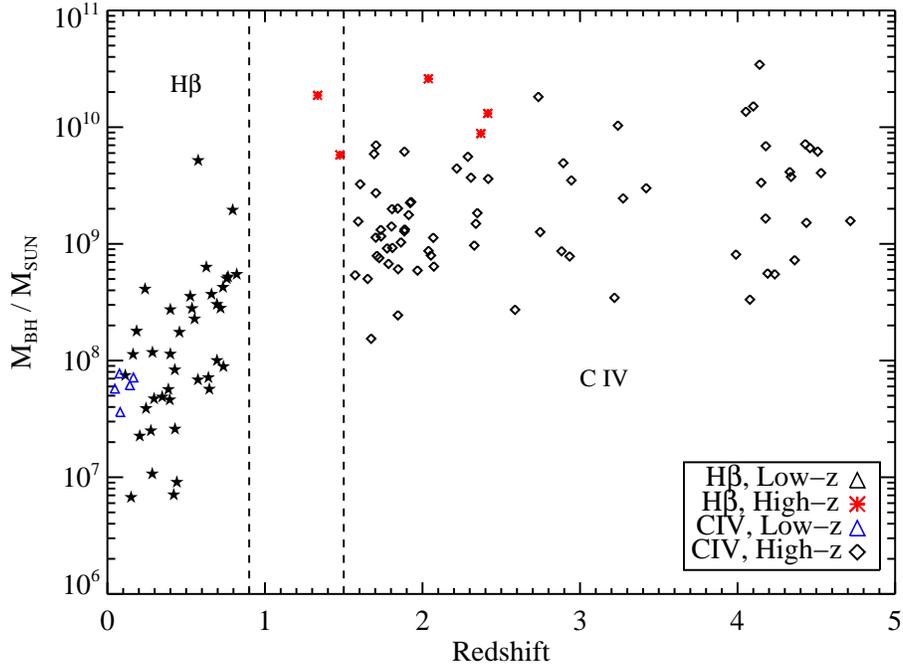}}}
    \caption{Distribution of estimated $M_{BH}$ as a function of $z$
    for the H$\beta$ sample (stars) and the C IV sample (open
    diamonds). The H$\beta$ sample probes sources with lower $M_{BH}$
    and $z$, while the C IV sample probes sources with higher $M_{BH}$
    and $z$. To break the degeneracy between emission line, $M_{BH}$,
    and $z$, we have collected a sample of H$\beta$ test sources (red
    asterisks) at high $M_{BH}$ and $z$, and a sample of C IV test
    sources (open blue triangles) at low $M_{BH}$ and $z$.
    \label{f-mbh_vs_z2}}
    \end{center}
\end{figure}

The opposite correlations for H$\beta$ and C IV are intriguing but may
represent problems with the broad line mass estimates. In particular,
it is possible that the error in the broad line mass estimates is
correlated with $\Gamma_X$, but in opposite ways for H$\beta$ and C
IV. The most likely source of such a spurious correlation would be a
correlation between $\Gamma_X$ and the scatter about the $R$--$L$
relationship for H$\beta$ and C IV, respectively. For example, if one
were to systematically overestimate $R$ with increasing $\Gamma_X$ for
the C IV emitting region, then one would infer a larger $M_{BH}$ from
C IV, and thus one would infer a spurious correlation between $M_{BH}$
and $\Gamma_X$. However, the H$\beta$ line is only available at $z
\lesssim 0.8$ and the C IV line is only available at $z \gtrsim 1.6$,
and thus the change in sign for the $\Gamma_X$--$M_{BH}$ correlations
could be due to different spectral components shifting into or out of
the observable X-ray spectral region (0.3--7.0 keV). In Figure
\ref{f-mbh_vs_z2}, we show the distribution of $M_{BH}$ as a function
of $z$ for the H$\beta$ and C IV samples. As is clear from Figure
\ref{f-mbh_vs_z2}, the C IV line is probing sources with $M_{BH}
\gtrsim 3 \times10^8 M_{\odot}$, while the H$\beta$ line is probing
sources with $M_{BH} \lesssim 3 \times 10^8 M_{\odot}$. Therefore, the
change in sign for the $\Gamma_X$ correlations could also be due to
something more physically interesting, such as a change in the
structure of the corona that occurs at some critical black hole mass
or accretion rate.

We can test if the change in sign for the $\Gamma_X$ correlations is
the result of problems with the mass estimates for either H$\beta$ or
C IV, or if it is the result of the differences in $z$ and $M_{BH}$
probed by the two lines. While the $\Gamma_X$--$M_{BH}$ relationship
is weak compared to the Eddington ratio dependencies, we can use the
$\Gamma_X$--$M_{BH}$ relationship to test whether the non-monotonic
dependency of $\Gamma_X$ on Eddington ratio is a spurious result
caused by systematic difference in the broad line mass estimates. This
is because the Eddington ratios are inferred from the broad line mass
estimates, and any systematic differences between the Eddington ratio
inferred from H$\beta$ as compared to C IV should also manifest
themselves in the weaker $\Gamma_X$--$M_{BH}$ relationship. This is
true irregardless of whether the difference in slopes between the
H$\beta$ and C IV $\Gamma_X$--$M_{BH}$ relationships is `statistically
significant' or not.

We compiled five sources from the literature at $z > 1.3$ with
H$\beta$-based mass estimates of $M_{BH} > 5 \times 10^9
M_{\odot}$. In addition, we compiled five more sources from the
literature at $z < 0.2$ and C IV-based mass estimates of $M_{BH} <
10^8 M_{\odot}$. This `test sample' of 10 sources is listed in Table
\ref{t-test}. Then, we test whether the H$\beta$ test sources are
better described by the H$\beta$ regression or by the C IV regression,
and likewise for the C IV test sources. If the change in sign for the
$\Gamma_X$ correlations is due to problems with the broad line mass
estimates, then we would expect the H$\beta$-based mass estimates to
be better described by the H$\beta$ regression. However, if the change
in sign is due to the difference in redshift and $M_{BH}$ probed by
the two regressions, then we would expect the H$\beta$ test sources to
be better described by the C IV regression, as the H$\beta$ test
sources are at high-$z$ and have high-$M_{BH}$. A similar argument
applies to the C IV test sources, since they are at low-$z$ and have
low black hole masses.

\begin{figure}
  \begin{center}
    \scalebox{0.6}{\rotatebox{90}{\plotone{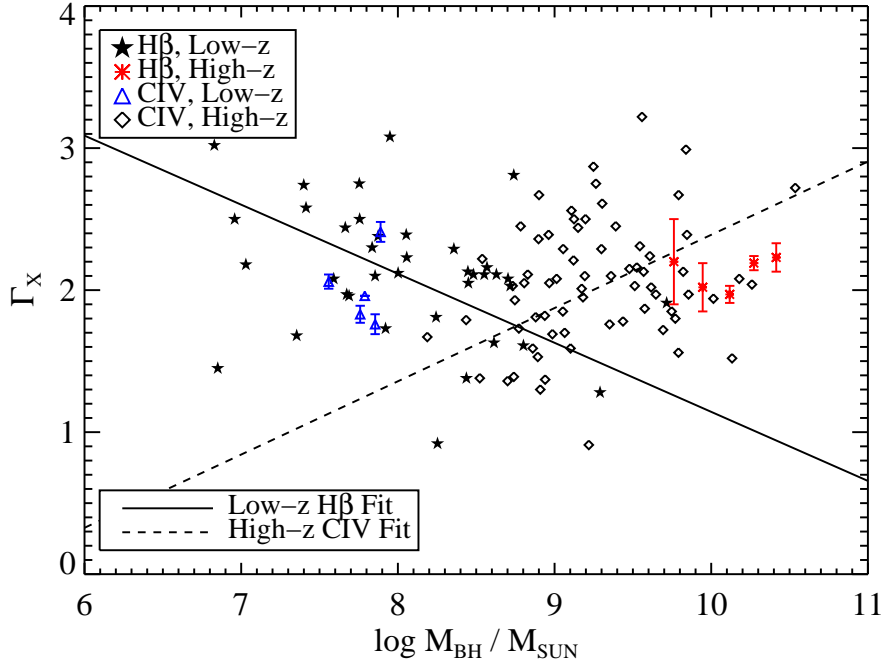}}}
    \caption{X-ray photon index as a function of estimated $M_{BH}$
    for low-$z$ sources with $M_{BH}$ derived from the H$\beta$ line,
    high-$z$ test sources with $M_{BH}$ derived from the H$\beta$
    line, low-$z$ test sources with $M_{BH}$ derived from the C IV
    line, and high-$z$ sources with $M_{BH}$ derived from the C IV
    line. The symbols are the same as in Figure
    \ref{f-mbh_vs_z2}. Also shown is the best fit regression using the
    H$\beta$ sample (solid line) and the C IV sample (dashed
    line). The high-$z$ H$\beta$ sources are better described by the
    high-$z$ C IV regression, and the low-$z$ C IV sources are better
    described by the low-$z$ H$\beta$ regression, implying that the
    difference in slopes between the H$\beta$ and C IV samples is not
    due to systematic differences in mass estimates derived from
    either line.
    \label{f-test1}}
  \end{center}
\end{figure}

Figure \ref{f-test1} compares $\Gamma_X$ and $M_{BH}$ for H$\beta$ and
C IV for both the sources in our main sample and the test sources, as
well as the best fit regression lines for the H$\beta$ and C IV
samples, respectively. The high-$z$, high-$M_{BH}$ H$\beta$ test
sources appear to be better described by the high-$z$, high-$M_{BH}$ C
IV-based regression, and likewise the low-$z$, low-$M_{BH}$ C IV test
sources appear to be better described by the low-$z$, low-$M_{BH}$
H$\beta$-based regression.

We can quantify this result by calculating the probability that the
H$\beta$ test sources `belong' to the H$\beta$-based regression, as
compared to the probability that the H$\beta$ test sources `belong' to
the C IV-based regression. Assuming that the test sources are as
equally likely to belong to either regression \emph{a priori}, this
ratio of probabilities is simply the ratio of the likelihood functions
of the test sources for each regression relationship, where the
likelihood functions are given by Equation (24) in \citet{kelly07a};
this ratio is called the `Bayes Factor' \citep[e.g.,][]{congdon06}. In
order to incorporate our uncertainty in the regression parameters, we
use the value of the likelihood function averaged over the probability
distribution of the regression parameters. We find that the H$\beta$
test sources are $\approx 250$ times more likely to `belong' to the C
IV-based regression, and that the C IV test sources are $\approx 140$
times more likely to `belong' to the H$\beta$-based
regression. Because the test sources are independent, it follows that
the test sources are $\gtrsim 10^4$ times more likely to be described
by the regression fit using the opposite emission line sample. This is
further evidence that $\Gamma_X$ and $M_{BH}$ are not statistically
independent; if $\Gamma_X$ and $M_{BH}$ were independent, then the
test sources would not show a strong preference for either
regression. Based on this analysis, we conclude that the change in
sign of the $\Gamma_X$ correlations is not due to problems associated
with the use of the H$\beta$ and C IV emission lines, but rather due
to the different range of $z$ and $M_{BH}$ probed by the two
subsamples.

\subsection{Is the Sign Change in the Correlations Caused by the Different 
  Redshift Ranges Probed?}

\label{s-ztest}

While it appears that $\Gamma_X$ has a nonmonotonic dependence on
$M_{BH}, L_{UV} / L_{Edd},$ and $L_X / L_{Edd}$, it is unclear as to
whether the sign change in the correlations is dependent on $z$ or
$M_{BH}$. However, we can test this in the same manner as was used to
test if the sign change is due to problems with the H$\beta$- or C
IV-based mass estimates. In this case, we need a sample of high-$z$,
low-$M_{BH}$ sources in order to break the degeneracy between $M_{BH}$
and $z$. If the sign change in the correlation is redshift dependent,
then we would expect the test sources to be better described by the
high-$z$, high-$M_{BH}$ C IV-based regression; but, if the sign change
is black hole mass dependent, then we would expect the test sources to
be better described by the low-$z$, low-$M_{BH}$ H$\beta$-based
regression.

\begin{figure}
  \begin{center}
    \scalebox{0.6}{\rotatebox{90}{\plotone{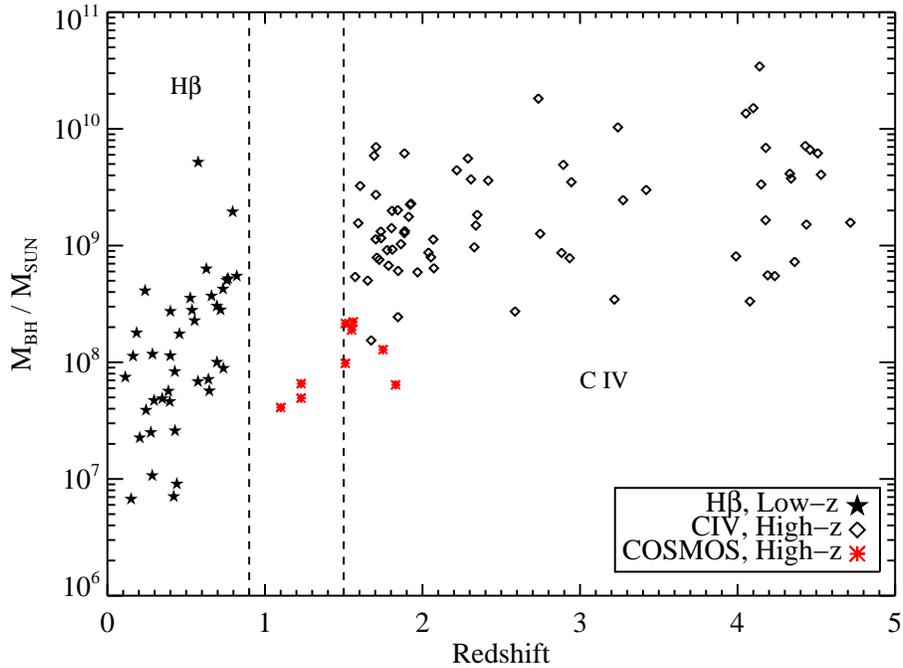}}}
    \caption{Same as Figure \ref{f-mbh_vs_z2}, but for the lower
    $M_{BH}$ and higher $z$ test sources from COSMOS. These test
    sources help break the degeneracy between the $M_{BH}$ and $z$
    present in our main sample.
    \label{f-mbh_vs_z_cosmos}}
    \end{center}
\end{figure}

\begin{figure}
  \begin{center}
    \scalebox{0.6}{\rotatebox{90}{\plotone{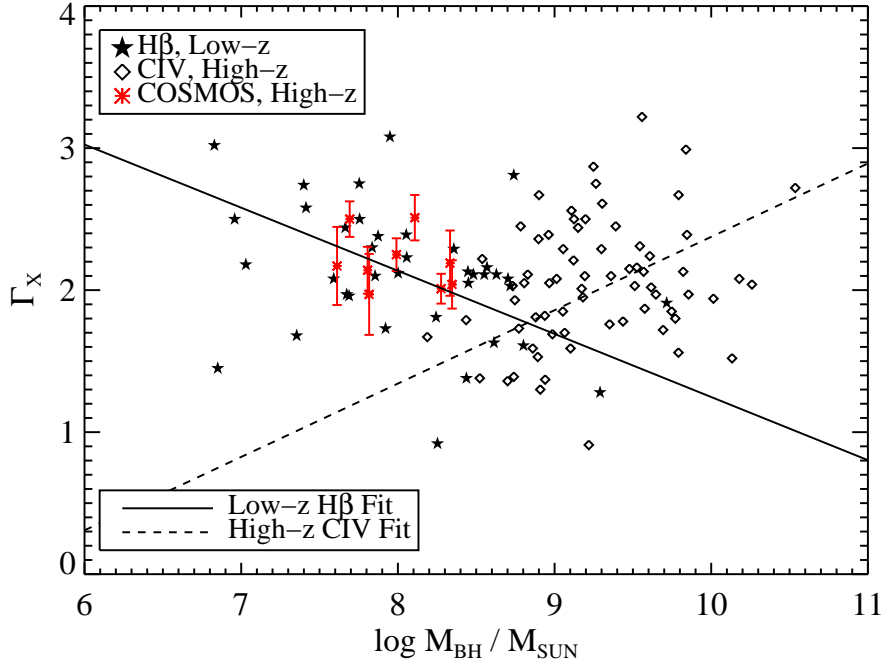}}}
    \caption{Same as Figure \ref{f-test1}, but for a sample of
    high-$z$, low-$M_{BH}$ test sources from \emph{COSMOS} (red
    asterisks with error bars). The \emph{COSMOS} sources are better
    described by the low-$z$, low-$M_{BH}$ regression, implying that
    the difference in slopes between the H$\beta$ and C IV samples is
    due to the difference in $M_{BH}$ probed by the two samples, and
    not due to the redshift differences.
    \label{f-test2}}
  \end{center}
\end{figure}

Our test sample consists of nine $z > 1, \hat{M}_{BL} < 3 \times 10^8
M_{\odot}$ quasars from the Cosmic Evolution Survey
\citep[COSMOS,][]{cosmos}, with optical spectra from Magellan
\citep{trump07} and X-ray spectra from XMM-Newton
\citep{mainieri07}. Black hole masses for these objects \citep[][in
preparation]{trump08} were estimated from the Mg II and C IV emission
lines in the same manner as above. The objects are summarized in Table
\ref{t-cosmos}, and their location in the $M_{BH}$--$z$ plane are
shown in Figure \ref{f-mbh_vs_z_cosmos}. The COSMOS sources break the
degeneracy between $M_{BH}$ and $z$ present in our SDSS sample, and
are therefore adequate to test for a redshift dependence in the slope
of the $\Gamma_X$--$M_{BH}$ relationship. In Figure \ref{f-test2} we
compare the COSMOS test sources with the H$\beta$- and C IV-based
regressions. As can be seen, the high-$z$, low-$M_{BH}$ COSMOS sources
are better described by the low-$z$, low-$M_{BH}$ regression. We can
quantify this in the same manner as described in \S~\ref{s-eline_test}
by averaging the likelihood function of the test sources over the
posterior probability distribution. We find that the COSMOS test
sources are $\gtrsim 10^5$ times more likely to be better described by
the H$\beta$-$\Gamma_X$ regression, and thus the sign change in the
$\Gamma_X$ correlations is not due to the difference in redshifts
probed by the H$\beta$ and C IV samples.

\section{DISCUSSION}

\label{s-discussion}

Previous work has found evidence for a correlation between
$\alpha_{\rm ox}$ and both $M_{BH}$ \citep{brunner97} and $L_{bol} /
L_{Edd}$ \citep{wang04}, and for a correlation between $\Gamma_X$ and
both $M_{BH}$ \citep[e.g.,][]{porquet04,picon05} and $L_{bol} /
L_{Edd}$ \citep[e.g.,][]{lu99,gier04,shemmer06}, in aggreement with
the results found in this work. However, our study differs from
previous work in that we study a large sample of RQQs (318 sources
with $\alpha_{\rm ox}$, 153 with $\Gamma_X$) over a broad range in
black hole mass ($10^6 \lesssim M_{BH} / M_{\odot} \lesssim 10^{10}$)
and redshift $0 < z < 4.8$; to date, this is the largest study of the
dependence of the X-ray properties of RQQs on $M_{BH}, L_{UV} /
L_{Edd},$ and $L_X / L_{Edd}$. In addition, this the first study of
its kind to correct for the intrinsic statistical uncertainty in broad
line mass estimates when quantifying the \emph{intrinsic} trends
between the X-ray emission and $M_{BH}, L_{UV} / L_{Edd},$ and $L_X /
L_{Edd}$.

Currently, there are two main types of geometries being considered for
the comptonizing corona. The first of these is that of a `slab'-type
geometry, possibly patchy, that sandwiches the disk
\citep[e.g.,][]{bis77,gal79,nayak00,sob04b}, and the second is that of
a hot spherical inner advection dominated flow
\citep[e.g.,][]{shap76,zdz99}; hybrids between the two geometries have
also been considered \citep[e.g.,][]{pout97,sob04a}. There is a
growing body of evidence that the advection dominated hot inner flow
does not exist in objects with Eddington ratios $L_{bol} / L_{Edd}
\gtrsim 0.01$, as inferred from the existence of a relativistically
broadened iron line \citep[e.g.,][]{mineo00,lee02,fabian02},
relativistically broadened reflection of ionized material
\citep{janiuk01}, and by analogy with galactic black holes
\citep[e.g.,][]{esin97,nowak02}. The range in Eddington ratios probed
by our study is at most $0.03 \lesssim L_{bol} / L_{Edd} \lesssim 2$,
with a mean of $L_{bol} / L_{Edd} \sim 0.25$. Therefore, the RQQs in
our study are likely to have disks that extend approximately down to
the last marginally stable orbit, and thus should only have the `slab'
type geometries.

\subsection{Dependence of $\alpha_{\rm ox}$ on $M_{BH}$}

\label{s-alfox_mbh}

In this work we have found that RQQs become more X-ray quiet as
$M_{BH}$ increases, and confirmed the well-established relationship
between $\alpha_{\rm ox}$ and $L_{UV}$. Because $L_{UV}$ increases
with $M_{BH}$ and the accretion rate relative to Eddington, $\dot{m}$,
the well-known $\alpha_{\rm ox}$--$L_{UV}$ correlation is likely
driven by the $\alpha_{\rm ox}$--$M_{BH}$ and $\alpha_{\rm
ox}$--$\dot{m}$ correlations. A correlation between $\alpha_{\rm ox}$
and $M_{BH}$ is expected even if the fraction of the bolometric
luminosity emitted by the disk is independent of $M_{BH}$, as the
effective temperature of the disk depends on $M_{BH}$. As $M_{BH}$
increases, the effective temperature of the disk decreases, thus
shifting the peak of the disk emission toward longer
wavelengths. Because the flux density at $2500$\AA\ lies redward of
the peak in the disk SED over most of the range $M_{BH}$ probed by our
study, this shift in the disk SED toward longer wavelengths produces
an increase in $L_{UV}$ relative to $L_X$.

We can use the standard thin disk solution to assess the evidence that
the fraction of energy emitted by the corona depends on $M_{BH}$. We
assume a simple model where the spectrum for the disk emission is that
expected for an extended thin accretion disk, and the spectrum for the
corona emission is a simple power-law with exponential cutoffs at the
low and high energy end. According to \citet{wandel00}, the spectrum
from a radially extended thin accretion disk can be approximated as
\begin{equation}
  f^{D}_{\nu} \approx A_{D} \left(\frac{\nu}{\nu_{co}}\right)^{-1/3} e^{-\nu / \nu_{co}}, 
  \label{eq-thindisk}
\end{equation}
where $A_{D}$ is the normalization and $\nu_{co}$ is the cut-off
frequency. In this work, we choose the normalization to ensure that
Equation (\ref{eq-thindisk}) integrates to unity, and therefore
$f^D_{\nu}$ gives the shape of the disk emission. For a Kerr black
hole, \citet{malkan91} finds that the cut-off frequency is related to
$M_{BH}$ as
\begin{equation}
h \nu_{co} = (6 {\rm eV}) \dot{m}^{1/4} (M_{BH} / 10^8 M_{\odot})^{-1/4}.
\label{eq-cutoff}
\end{equation}
We assume that the X-ray emission from the corona can be described by
a simple power law with an exponential cutoff at the high and low end:
\begin{equation}
  f^{C}_{\nu} = A_C \nu^{-(\Gamma_X - 1)} e^{-\nu / \nu_{high}} e^{-\nu_{low} / \nu}.
  \label{eq-corona}
\end{equation}
Here, $A_C$ is the corona spectrum normalization, $\nu_{high}$ is the
high energy cutoff, and $\nu_{low}$ is the low energy cutoff. We
choose the low energy cutoff to be $\nu_{low} = 20\ {\rm eV}$, and we
choose the high energy cutoff to be $\nu_{high} = 200\ {\rm keV}$
\citep[e.g.,][]{gilli07}. As with Equation (\ref{eq-thindisk}), we
choose the normalization in Equation (\ref{eq-corona}) to be equal to
unity.

Denoting $f_D$ to be the fraction of bolometric luminosity emitted by
the disk, our model RQQ spectrum is then
\begin{equation}
  L_{\nu} \approx L_{bol} \left[f_D f^D_{\nu} + (1 - f_D) f^C_{\nu} \right].
  \label{eq-modelspec}
\end{equation}
We computed Equation (\ref{eq-modelspec}) assuming a value of $\dot{m}
= 0.2$ and $f_D = 0.85$. We chose the value of the $\dot{m} = 0.2$
because it is representative of the RQQs in our sample, and we chose
the value $f_D = 0.85$ because it gives values of $\alpha_{\rm ox}$
typical of the RQQs in our sample. We vary $M_{BH}$ but keep $f_D$ and
$\dot{m}$ constant because we are interested in investigating whether
there is evidence that assuming independence between $M_{BH}$ and both
$f_D$ and $\dot{m}$ is inconsistent with our $\alpha_{\rm ox}$
results.

\begin{figure}
  \begin{center}
    \includegraphics[scale=0.33,angle=90]{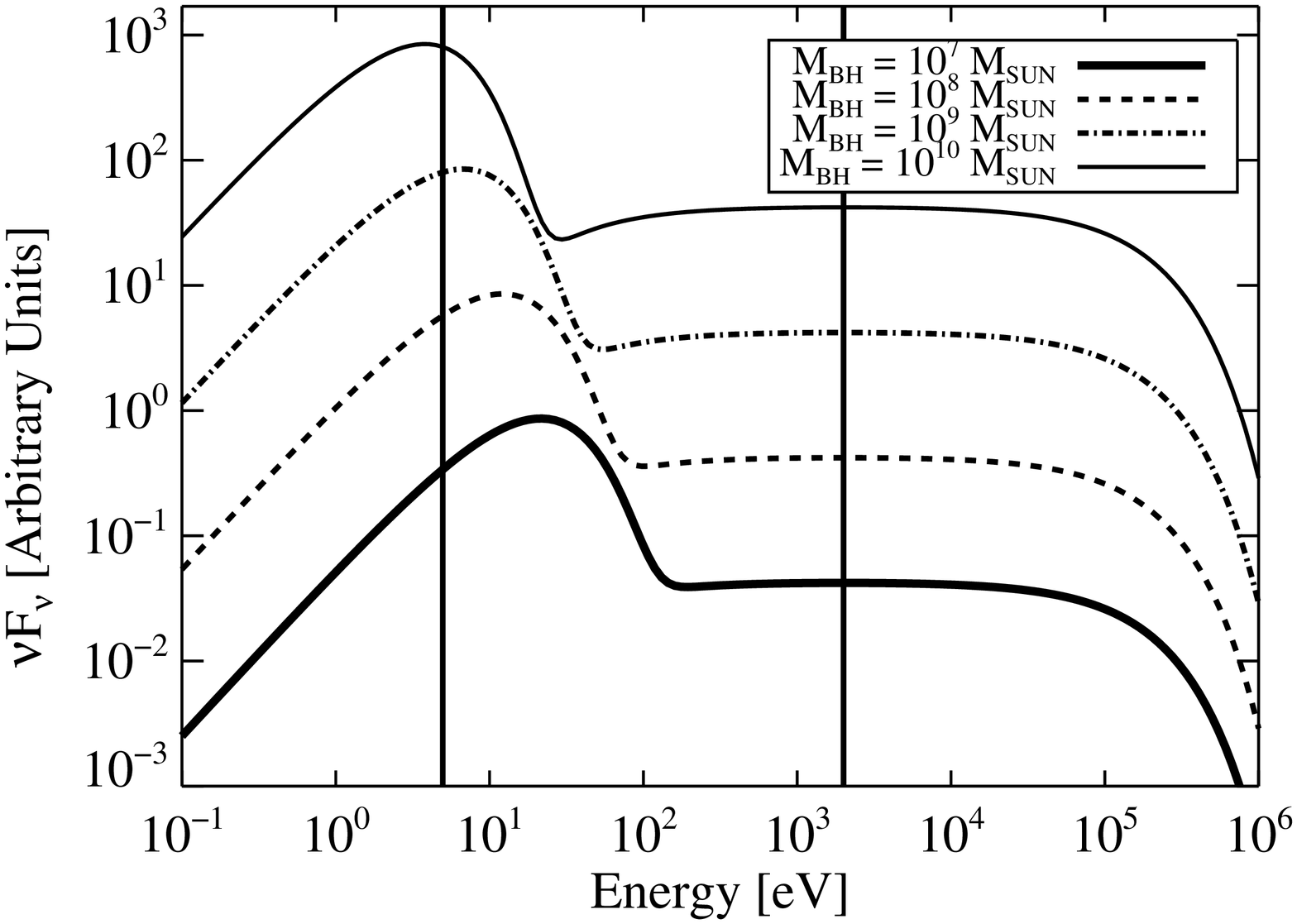}
    \includegraphics[scale=0.33,angle=90]{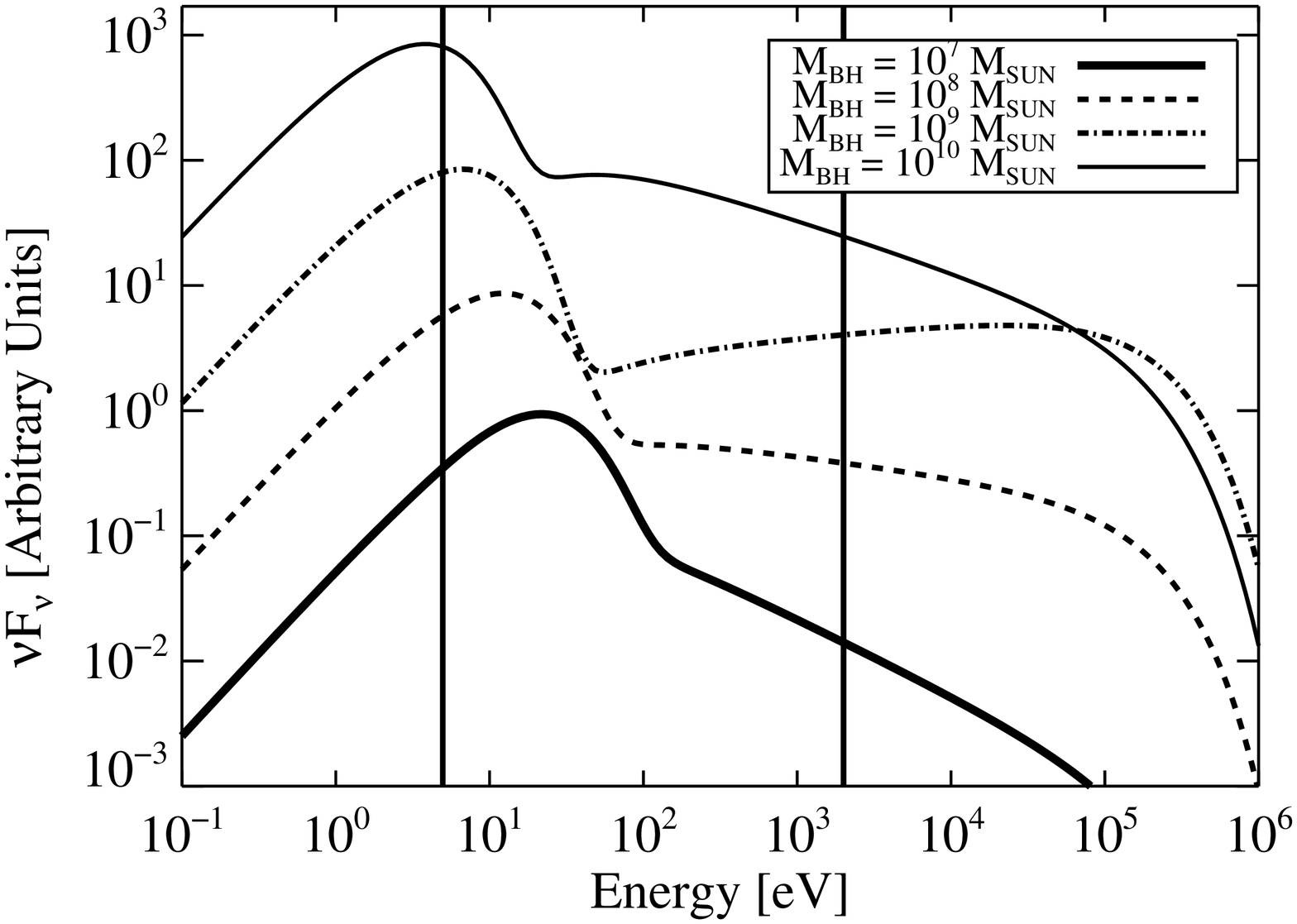}
    \caption{Model RQQ spectra computed from Equations
    (\ref{eq-thindisk})--(\ref{eq-modelspec}), assuming $\Gamma_X = 2$
    (left) and a varying $\Gamma_X$ with $M_{BH}$ (right). The spectra
    are computed for a RQQ with $M_{BH} / M_{\odot} = 10^7$ (thick
    solid line), $10^8$ (dashed line), $10^9$ (dashed-dotted line),
    and $10^{10}$ (thin solid line). In all cases we assume $\dot{m} =
    0.2$ and that $f_D = 85\%$ of the bolometric luminosity is emitted
    by the disk. The vertical lines mark the locations of 2500\AA\
    and 2 keV. The dependence of the location of the peak in the disk
    emission on $M_{BH}$ is apparent, producing a correlation between
    $\alpha_{\rm ox}$ and $M_{BH}$ even if the fraction of bolometric
    luminosity emitted by the disk is independent of $M_{BH}$.
    \label{f-modelspec}}.
  \end{center}
\end{figure}

We compute Equation (\ref{eq-modelspec}) for two forms of the
dependence of $\Gamma_X$ on $M_{BH}$. For the first model, we assume a
constant value of $\Gamma_X = 2$. For the second model, we assume that
$\Gamma_X$ depends on $M_{BH}$ according to our best fit regression
results, where $\Gamma_X$ depends on $M_{BH}$ according to Equation
(\ref{eq-gamxmbh_hbeta}) for $M_{BH} \lesssim 3 \times 10^8
M_{\odot}$, and $\Gamma_X$ depends on $M_{BH}$ according to Equation
(\ref{eq-gamxmbh_civ}) for $M_{BH} \gtrsim 3 \times 10^8
M_{\odot}$. We ignore the intrinsic dispersion in $\Gamma_X$. In
Figure \ref{f-modelspec} we show the spectra computed from Equation
(\ref{eq-modelspec}) for RQQs with $M_{BH} / M_{\odot} = 10^7, 10^8,
10^9,$ and $10^{10}$. The dependence of the location of the peak in
the disk emission on $M_{BH}$ is clearly illustrated.

\begin{figure}
  \begin{center}
    \scalebox{0.6}{\rotatebox{90}{\plotone{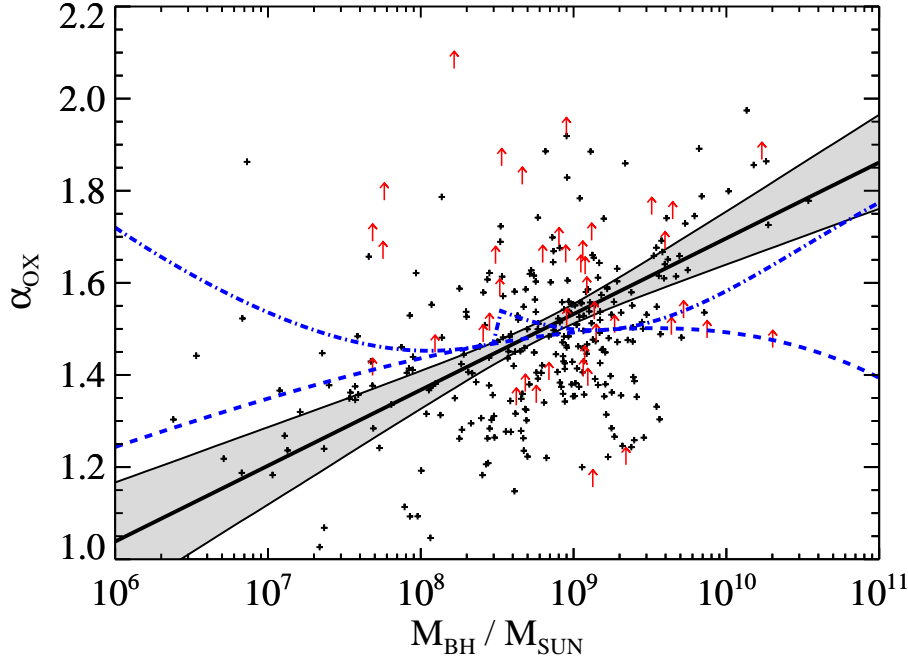}}}
    \caption{Dependence of $\alpha_{\rm ox}$ on $M_{BH}$ computed from
    Equations (\ref{eq-thindisk})--(\ref{eq-modelspec}), assuming
    $\Gamma_X = 2$ (blue dashed line) and a varying $\Gamma_X$ with
    $M_{BH}$ (blue dot-dashed line). As with Figure \ref{f-modelspec},
    we compute Equations (\ref{eq-thindisk})--(\ref{eq-modelspec})
    assuming $\dot{m} = 0.2$ and $f_D = 0.85$. The predictions from
    the model RQQ spectra are compared with our observed data and the
    regression results, where the symbols and lines have the same
    meaning as in Figure \ref{f-alfox_reg}. The $\alpha_{\rm
    ox}$--$M_{BH}$ relationships predicted from assuming that $f_D$ is
    independent of $M_{BH}$ are inconsistent with the observed
    $\alpha_{\rm ox}$--$M_{BH}$ relationship.
    \label{f-alfox_model}}
  \end{center}
\end{figure}

In Figure \ref{f-alfox_model} we compare the $\alpha_{\rm
ox}$--$M_{BH}$ regression results with the dependence of $\alpha_{\rm
ox}$ on $M_{BH}$ expected from Equation (\ref{eq-modelspec}) for both
$\Gamma_X$--$M_{BH}$ models. Under the thin disk approximation, a
correlation is expected between $\alpha_{\rm ox}$ and $M_{BH}$, even
if the fraction of bolometric luminosity emitted by the disk is
independent of $M_{BH}$. However, our data are inconsistent with the
assumption that $f_D$ and $M_{BH}$ are independent, given the thin
disk approximation. Under the assumption that $f_D$ and $M_{BH}$ are
independent, the $\alpha_{\rm ox}$--$M_{BH}$ correlation is too flat,
and a increase in the fraction of bolometric luminosity emitted by the
disk with increasing $M_{BH}$ is needed to match the steeper observed
dependence of $\alpha_{\rm ox}$ on $M_{BH}$. Alternatively, if $f_D$
increases with increasing $\dot{m}$, as we argue in
\S~\ref{s-alfox_mdot}, then a steeper $\alpha_{\rm ox}$--$M_{BH}$
correlation also results if $M_{BH}$ and $\dot{m}$ are correlated. In
this case, if $f_D$ increases with increasing $\dot{m}$, and if
$M_{BH}$ increases with increasing $\dot{m}$, then $f_D$ will also
increase with increasing $M_{BH}$, thus producing a steeper observed
$\alpha_{\rm ox}$--$M_{BH}$ correlation.

To the extant that Equations (\ref{eq-thindisk})--(\ref{eq-modelspec})
accurately approximate the spectral shape of RQQs, our data imply that
either the fraction of bolometric luminosity emitted by the disk
increases with increasing $M_{BH}$, that $M_{BH}$ and $\dot{m}$ are
correlated, or both. Some theoretical models have suggested that the
fraction of bolometric luminosity emitted by the disk should depend on
$\dot{m}$, but be relatively insensitive to $M_{BH}$
\citep[e.g.,][]{czerny03,liu03}. Therefore, while a significant
dependence of $f_D$ on $M_{BH}$ is not predicted by these disk/corona
models, these models are still consistent with the interpretation that
a $M_{BH}$--$\dot{m}$ correlation is driving the steeper $\alpha_{\rm
ox}$--$M_{BH}$ correlation. Unfortunately, without accurate estimates
of $\dot{m}$ we are unable to distinguish between these two
possibilities.

A shift in the peak of the disk SED with $M_{BH}$ may also explain the
dependence of $\alpha_{\rm ox}$ on redshift observed by K07. K07
speculated that the observed hardening of $\alpha_{\rm ox}$ with
increasing $z$ at a given $L_{UV}$ may be due to a correlation between
$\alpha_{\rm ox}$ and $M_{BH}$, manifested through a $M_{BH}$--$z$
correlation. At a given $L_{UV}$, an increase in $M_{BH}$ will result
in an increase in $L_X$ relative to $L_{UV}$, assuming that $\dot{m}$
is not strongly correlated with $M_{BH}$. This is because an increase
in $M_{BH}$ decreases the temperature of the disk, shifting the peak
in the disk SED toward the red, and thus increasing the luminosity at
2500\AA. However, since K07 investigated the dependence of
$\alpha_{\rm ox}$ on $z$ at a given $L_{UV}$, the luminosity at
2500\AA\ is held constant. Therefore, the overall disk emission must
decrease in order to keep the luminosity at 2500\AA\ constant despite
the increase in $M_{BH}$. As a result, an increase in $M_{BH}$ at a
given $L_{UV}$ will result in an increase in the X-ray luminosity
relative to the luminosity at 2500\AA. Because $M_{BH}$ and $z$ are
correlated in our flux limited sample (e.g., see Figure
\ref{f-mbh_vs_z}), an increase in $z$ will probe RQQs with higher
$M_{BH}$. As a result, RQQs will become more X-ray loud with
increasing $z$, at a given 2500\AA\ luminosity. Consequently, deeper
surveys that probe a greater range of $M_{BH}$ should not see as
strong of a correlation between $M_{BH}$ and $z$, thereby reducing the
magnitude of a $\alpha_{\rm ox}$--$z$ correlation. Indeed,
investigations based on samples that span a greater range in
luminosity do not find evidence for a correlation between $\alpha_{\rm
ox}$ and $z$ \citep[e.g.,][]{steffen06,just07}, qualitatively
consistent with our interpretation of a $\alpha_{\rm ox}$--$z$
correlation.

\subsection{Dependence of $\alpha_{\rm ox}$ on $\dot{m}$}

\label{s-alfox_mdot}

We have found that $\alpha_{\rm ox}$ increases with increasing $L_{UV}
/ L_{Edd}$, and decreases with increasing $L_X / L_{Edd}$. The mere
existence of these correlations is not particularly interesting, as we
would expect that the ratio of optical/UV luminosity to X-ray
luminosity would increase as the fraction of optical/UV luminosity
relative to Eddington increases, and vice versa for an increase in
$L_X / L_{Edd}$. However, the relative magnitude of these dependencies
carries some information regarding the dependence of $\alpha_{\rm ox}$
on $\dot{m}$. A correlation between $\alpha_{\rm ox}$ and $L_{UV} /
L_{Edd}$ implies that $L_{UV} / L_X$ increases as the quantity $f_{UV}
\dot{m}$ increases, where $f_{UV}$ is the fraction of bolometric
luminosity emitted at 2500\AA. Likewise, an anti-correlation between
$\alpha_{\rm ox}$ and $L_X / L_{Edd}$ implies that $L_{UV} / L_X$
decreases as the quantity $f_X \dot{m}$ increases, where $f_X$ is the
fraction of bolometric luminosity emitted at 2 keV. If the fraction of
the bolometric luminosity emitted by the disk, as compared to the
corona, increases with increasing $\dot{m}$, then we would expect a
strong increase in $L_{UV} / L_X$ with the product $f_{UV} \dot{m}$,
resulting from the dual dependency of $L_{UV} / L_X$ on $f_{UV}$ and
$\dot{m}$. Furthermore, because the fraction of bolometric luminosity
emitted by the disk should decrease with increasing $f_X$, then, if
the fraction of bolometric luminosity emitted by the disk increases
with increasing $\dot{m}$, we would expect a weaker dependence of
$L_{UV} / L_X$ on the quantity $f_X \dot{m}$. This is because an
increase in $\dot{m}$ causes an increase in the disk emission relative
to the corona emission, which will then work against the decrease in
disk emission relative to the corona that results from an increase in
$f_X$. The end result is a weaker dependence of $L_{UV} / L_X$ on the
product $f_X \dot{m}$. Indeed, this is what we observe, where $L_{UV}
/ L_X \propto (L_{UV} / L_{Edd})^{2.5}$ and $L_{UV} / L_X \propto (L_X
/ L_{Edd})^{-1.5}$. Therefore, we conclude that the disk emission
relative to the corona emission increases with increasing
$\dot{m}$. This is in agreement with some models of corona with a slab
geometry \citep[e.g.,][]{czerny97,janiuk00,merloni02,liu03}, where the
$\alpha_{\rm ox}$--$\dot{m}$ correlation arises due to a dependency of
the size of the corona on $\dot{m}$.

Our result that $\alpha_{\rm ox}$ is correlated with $L_{UV} /
L_{Edd}$ and anti-correlated with $L_X / L_{Edd}$ is inconsistent with
a constant bolometric correction to both the optical/UV and X-ray
luminosities. Instead, an increase in $L_{UV} / L_X$ with increasing
$\dot{m}$ implies that the bolometric correction depends on
$\dot{m}$. Because we conclude that the fraction of bolometric
luminosity emitted by the disk increases with increasing $\dot{m}$,
this implies that the bolometric correction to the optical/UV
luminosity decreases with increasing $\dot{m}$, while the bolometric
correction to the X-ray luminosity increases with increasing
$\dot{m}$. The direction of this trend is consistent with the results
of \citet{vasud07}, who find that the bolometric correction to the
X-ray luminosity increases with increasing $L_{bol} /
L_{Edd}$. Similarly, we have found evidence that the fraction of
bolometric luminosity emitted by the disk depends on $M_{BH}$,
therefore implying that the bolometric correction also depends on
$M_{BH}$. Even if the fraction of bolometric luminosity emitted by the
disk is independent of $M_{BH}$, the bolometric correction will still
depend on $M_{BH}$ because the location of the peak in the disk
emission will shift toward longer wavelengths as $M_{BH}$
increases. As $M_{BH}$ varies, the luminosity at 2500\AA\ probes a
different region of the quasi-blackbody disk emission, thereby
producing a dependence of bolometric correction on $M_{BH}$.

\subsection{Implications for Black Hole Feedback}

A significant amount of recent work suggests that radiative and
mechanical feedback energy from AGN plays an important part in galaxy
and supermassive black hole coevolution
\citep[e.g.,][]{fabian99,wyithe03,hopkins05}. Within the context of
these models, a nuclear inflow of gas, possibly the result of a galaxy
merger, feeds the SMBH, thus igniting a quasar. The SMBH grows until
feedback energy from the quasar is able to drive out the accreting
gas, thus halting the accretion process. Hydrodynamic calculations of
accretion flows have shown that the efficiency of the quasar in
driving an outflow depends on the fraction of energy emitted through
he UV/disk component as compared to the X-ray/corona component
\citep{proga07}. The disk component produces luminosity in the UV,
which is responsible for driving an outflow via radiation pressure on
lines, whereas the corona component produces luminosity in the X-rays,
which is responsible for driving an outflow via thermal
expansion. Calculations by \citet{proga07} have shown that radiation
driving produces an outflow that carries more mass and energy than
thermal driving. If the efficiency of black hole feedback depends on
the quasar SED, any dependence on $M_{BH}$ and $\dot{m}$ of the
fraction of AGN energy emitted in the UV as compared to the X-ray has
important consequences for models of black hole growth.

Because we have found evidence that the fraction of bolometric
luminosity emitted by the disk increases with increasing $\dot{m}$ and
$M_{BH}$, this implies that black holes becomes more efficient at
driving an outflow with increasing $\dot{m}$ and $M_{BH}$. However,
the $\alpha_{\rm ox}$--$M_{BH}$ correlation may be due to the
combination of both a correlation between $M_{BH}$ and $\dot{m}$, and
a dependence of the location peak in the disk SED on $M_{BH}$. If the
fraction of energy emitted by the disk only depends weakly on
$M_{BH}$, as some theoretical models have suggested
\citep[e.g.,][]{czerny03,liu03}, the fraction of energy emitted in the
UV will still decrease with increasing $M_{BH}$ becuase the peak of
the disk emission will shift away from the UV. In this case, at a
given $\dot{m}$ we would expect that black holes will become less
efficient at driving an outflow with increasing $M_{BH}$.

\subsection{Dependence of $\Gamma_X$ on $M_{BH}$ and $\dot{m}$}

In this work we have also found evidence that $\Gamma_X$ and $M_{BH},
L_{UV} / L_{Edd},$ and $L_X / L_{Edd}$ are not statistically
independent. Moreover, the dependence of $\Gamma_X$ on black hole mass
or Eddington ratio appears to follow a non-monotonic form, although
the $\Gamma_X$--$M_{BH}$ trend is weak compared to the dependency of
$\Gamma_X$ on Eddington ratio. For the $\Gamma_X$--$M_{BH}$
relationship, the X-ray continuum hardens with increasing black hole
mass until $M_{BH} \sim 3 \times 10^8 M_{\odot}$, after which the
X-ray continuum softens with increasing black hole mass. The opposite
is true for the $\Gamma_X$--$L_{UV} / L_{Edd}$ and $\Gamma_X$--$L_X /
L_{Edd}$ trends, and further work is needed to confirm this
result. Previous studies have not seen this non-monotonic trend
because they have only employed the H$\beta$ emission line, and
therefore their samples have been dominated by low-$z$, low-$M_{BH}$
sources.

\subsubsection{Selection Effects}

\label{s-seleff}

It is unlikely that the dependence of $\Gamma_X$ on $M_{BH}, L_{UV} /
L_{Edd},$ or $L_X / L_{Edd}$ is due to redshifting of the observable
spectra range. If this were the case, then as $M_{BH}$ increases, so
does $z$ due to selection effects, and thus we would observe a
decrease in $\Gamma_X$ as the `soft excess' shifts out of the observed
0.3--7 keV spectral range, while the compton reflection component
shifts into the observed spectral range. However, there are lines of
evidence that suggest that the $\Gamma_X$ correlation are not due to
redshifting of the observable spectral region, and that at least some
of the observed dependency of $\Gamma_X$ on $M_{BH}, L_{UV} /
L_{Edd},$ and $L_X / L_{Edd}$ is real. First, in \S~\ref{s-ztest} we
tested whether a sample of nine $z > 1$ test sources with $M_{BH}
\lesssim 3 \times 10^8 M_{\odot}$ were better described by a
regression fit using the $z > 1.5, M_{BH} \gtrsim 3 \times 10^8
M_{\odot}$ sources, or by a regression fit using the $z < 1, M_{BH}
\lesssim 3 \times 10^8 M_{\odot}$ sources. We found that the test
sources were better fit using the regression of similar $M_{BH}$, and
therefore that the difference in the $\Gamma_X$--$M_{BH}$ correlations
primarily depends on $M_{BH}$. Second, similar trends at low redshift
between $\Gamma_X$ and $M_{BH}$ or $L_{bol} / L_{Edd}$ have been seen
in other studies that only analyze the hard X-ray spectral slope
\citep[typically 2--12 keV, e.g.,][]{picon05,shemmer06}, and thus
these studies are not effected by the soft excess. Third, the compton
reflection hump is unlikely to shift into the observabable spectral
range until $z \sim 1$. However, the contribution to the inferred
$\Gamma_X$ from compton reflection at $z \gtrsim 1$ is likely weak, if
not negligible, as our $z \gtrsim 1$ sources have $M_{BH} \gtrsim 10^8
M_{\odot}$ and are highly luminous, and therefore are expected to only
have weak reflection components \citep{mineo00,ball01,bianchi07}.

There are two scenarios in which the non-monotonic behavior of
$\Gamma_X$ with $M_{BH}$ or Eddington ratio may be artificially caused
by selection. We will focus on the Eddington ratio dependency, as it
is the strongest; however, our argument also applies to
$M_{BH}$. First, the intrinsic dependency of $\Gamma_X$ on Eddington
ratio could be linear with increasing intrinsic scatter at high
$L_{bol} / L_{Edd}$. Then, an inferred non-monotonic trend would occur
if we were to systematically miss quasars with high $L_{bol} /
L_{Edd}$ and steep X-ray spectra. Alternatively, there could be no
intrinsic dependency of $\Gamma_X$ on Eddington ratio. In this case,
we would infer a non-monotonic trend if we were to systematically miss
quasars with steep X-ray spectra at low and high $L_{bol} / L_{Edd}$,
and quasars with flat X-ray spectra at moderate $L_{bol} / L_{Edd}$.

We do not consider it likely that the observed non-monotonic
dependence of $\Gamma_X$ on Eddington ratio is due solely to selection
effects. K07 describes the sample selection for sources with
$\Gamma_X$. With the exception of some of the $z > 4$ quasars, all
sources from K07 were selected by cross-correlating the SDSS DR3
quasars with public \emph{Chandra} observations. Almost all SDSS
sources in K07 had serendipitious \emph{Chandra} observations, and
therefore were selected without regard to their X-ray properties. K07
estimated $\Gamma_X$ for all sources that were detected in X-rays at
the level of $3\sigma$ or higher. Therefore, the only additional
criterion beyond the SDSS selection imposed by K07 is the requirement
that the source had to be detected in X-ray, which was fulfilled by
$90\%$ of the quasars; the undetected sources were slightly more
likely to be found at lower redshift, probably due to the presence of
unidentified BAL quasars. As a result, the K07 sample selection
function is essentially equivalent to the SDSS quasar selection
function. Because the SDSS selects quasars based on their optical
colors, the most likely cause of selection effects is the optical
color selection. There is evidence that $\Gamma_X$ is correlated with
the slope of the optical continuum, where the X-ray continuum flattens
(hardens) as the optical continuum steepens (softens)
\citep[][K07]{gall05}. The SDSS selection probability is lower for red
sources \citep{rich06}, so we might expect to systematically miss
sources with smaller $\Gamma_X$. However, for the two scenarios
described above, this is opposite the trend needed to explain the
$\Gamma_X$--$L_X / L_{Edd}$ relationship, where we need to at least
systematically miss sources with larger $\Gamma_X$. Furthermore, the
drop in SDSS selection efficiency with optical spectral slope only
occurs at $2 < z < 4$ \citep{rich06}, thus we would expect a redshift
dependence for this selection effect. As we have argued above, and in
\S~\ref{s-ztest}, the non-monotonic trends for $\Gamma_X$ cannot be
completely explained as the result of different redshift ranges being
probed.

\subsubsection{Implications for Accretion Physics}

\label{s-accphys}

The dependence of $\Gamma_X$ on $L_{UV} / L_{Edd}$ and $L_X / L_{Edd}$
is likely due to a dependence of $\Gamma_X$ on $\dot{m}$. If these
$\Gamma_X$ correlations were due to a dependence of $\Gamma_X$ on
$f_{UV}$ or $f_X$, then we would expect opposite trends for $L_{UV} /
L_{Edd}$ and $L_X / L_{Edd}$, as $f_{UV}$ and $f_X$ should be
anti-correlated. The fact that the regression results for the
$\Gamma_X$--$L_{UV} / L_{Edd}$ and $\Gamma_X$--$L_X / L_{Edd}$
relationships are similar implies that $\Gamma_X$ depends on
$\dot{m}$, and at most only weakly on $f_{UV}$ or $f_X$.

A non-monotonic dependence of $\Gamma_X$ on $\dot{m}$ is predicted
from the accreting corona model of \citet{janiuk00}, as well as a
non-monotonic dependence of $\Gamma_X$ on the viscosity
\citep{bech03}. In addition, $\Gamma_X$ is expected to steepen with
increasing optical depth
\citep[e.g.,][]{haardt91,haardt93,czerny03}. One could then speculate
that the dependence of $\Gamma_X$ on $M_{BH}$ or $\dot{m}$ is due to a
non-monotonic dependence of the corona optical depth on $\dot{m}$,
which may indicate a change in the structure of the disk/corona system
at $\sim 3 \times 10^8 M_{\odot}$ or some critical $\dot{m}$. Recent
work also suggests a non-monotonic dependence of the optical/UV
spectral slope, $\alpha_{UV}$, on $\dot{m}$
\citep{bonning07,davis07}. From this work, it has been inferred that
the optical/UV continuum becomes more red with increasing $\dot{m}$
until $L_{bol} / L_{Edd} \approx 0.3$, after which the optical/UV
continuum becomes more blue with increasing $\dot{m}$. Assuming the
bolometric corrections described in \S~\ref{s-eddrat}, the turnover in
the $\Gamma_X$--$L_{bol} / L_{Edd}$ relationship also occurs at
$L_{bol} / L_{Edd} \approx 0.3$. \citet{bonning07} suggested that the
turnover in spectral slope at $L_{bol} / L_{Edd} \sim 0.3$ may be due
to a change in accretion disk structure, where the inner part of the
accretion disk becomes thicker due to increased radiation pressure
\citep{abram88}. \citet{bonning07} performed a simple approximation to
this `slim disk' solution and found that it is able to produce a
non-monotonic trend between optical color and Eddington ration. If the
inner disk structure changes at high $\dot{m}$, this change could
alter the corona structure, producing the observed trend between
$\Gamma_X$ and Eddington ratio.

Unfortunately, current models for corona geometry make a number of
simplifying assumptions, and do not yet predict a specific
relationship between $\alpha_{\rm ox}, \Gamma_X, M_{BH},$ and
$\dot{m}$. Ideally, full magneto-hydrodynamic simulations
\citep[e.g.,][]{devill03,turner04,krolik05} that include accretion
disk winds \citep[e.g.,][]{murray95,proga04} should be used to
interpret the results found in this work. However, MHD simulations
have not advanced to the point where they predict the dependence of
$\alpha_{\rm ox}$ and $\Gamma_X$ on quasar fundamental parameters, but
hopefully recent progress in analytical descriptions of the
magneto-rotational instability \citep{pessah06,pessah07} will help to
overcome some of the computational difficulties and facilitate further
advancement.

\section{SUMMARY}

\label{s-summary}

In this work we have investigated the dependence of $\alpha_{\rm ox}$
and $\Gamma_X$ on black hole mass and Eddington ratio using a sample
of 318 radio-quiet quasars with X-ray data from \emph{ROSAT}
\citep{strat05} and \emph{Chandra} \citep{kelly07c}, and optical data
mostly from the SDSS; 153 of these sources have estimates of
$\Gamma_X$ from \emph{Chandra}. Our sample spans a broad range in
black hole mass ($10^6 \lesssim M_{BH} / M_{\odot} \lesssim 10^{10}$),
redshift ($0 < z < 4.8$), and luminosity ($10^{43} \lesssim \lambda
L_{\lambda} (2500$\AA$) [{\rm erg\ s^{-1}}] \lesssim 10^{48}$). To
date, this is the largest study of the dependence of RQQ X-ray
parameters on $M_{BH}, L_{UV} / L_{Edd},$ and $L_X / L_{Edd}$. Our
main results are summarized as follows:
\begin{itemize}
\item
  We show that $\alpha_{\rm ox}$ is correlated with $L_{UV} / L_{Edd}$
  and anti-correlated with $L_X / L_{Edd}$. This result is
  inconsistent with a constant bolometric correction being applicable
  to both the optical/UV luminosity and the X-ray luminosity. This
  result, when taken in combination with recent work by
  \citet{vasud07}, implies that constant bolometric corrections can be
  considerably unreliable and lead to biased results. Instead, we
  argue that $L_{UV} / L_X$ increases with increasing $\dot{m}$ and
  increasing $M_{BH}$, therefore implying that the bolometric
  correction depends on $\dot{m}$ and $M_{BH}$.
\item
  We performed a linear regression of $\alpha_{\rm ox}$ on luminosity,
  black hole mass, $L_{UV} / L_{Edd}$, and $L_X / L_{Edd}$, and found
  significant evidence that $\alpha_{\rm ox}$ depends on all four
  quantities: $L_{UV} / L_{X} \propto L_{UV}^{0.31 \pm 0.03}, L_{UV}
  / L_{X} \propto M_{BH}^{0.43 \pm 0.06}, L_{UV} / L_{X} \propto
  (L_{UV} / L_{Edd})^{2.57 \pm 0.45},$ and $L_{UV} / L_X \propto
  L_{X} / L_{Edd}^{-1.48 \pm 0.14}$. The dependence of $\alpha_{\rm
  ox}$ on $L_{UV}$ may be due to the dual dependence of $\alpha_{\rm
  ox}$ on $M_{BH}$ and $\dot{m}$. Because we have attempted to correct
  for the statistical uncertainties in $\alpha_{\rm ox}$ and the broad
  line estimates of $M_{BH}$, these results refer to the
  \emph{intrinsic} relationships involving $\alpha_{\rm ox}$ and
  $M_{BH}$, and are not merely the relationships between $\alpha_{\rm
  ox}$ and the broad line mass estimates, $\hat{M}_{BL} \propto
  L^{\gamma} FWHM^2$.
\item
  A correlation between $\alpha_{\rm ox}$ and $M_{BH}$ is expected
  from the fact that the peak in the disk emission will shift to
  longer wavelengths as $M_{BH}$ increases, even if the fraction of
  the bolometric luminosity emitted by the disk does not change with
  $M_{BH}$. Using a simple model for RQQ spectra, we argue that the
  observed $\alpha_{\rm ox}$--$M_{BH}$ correlation is steeper than
  that expected if both $\dot{m}$ and the fraction of bolometric
  luminosity produced by the disk are independent of $M_{BH}$. The
  observed $\alpha_{\rm ox}$--$M_{BH}$ relationship therefore implies
  that either the fraction of bolometric luminosity emitted by the
  disk increases with increasing $M_{BH}$, that $M_{BH}$ is correlated
  with $\dot{m}$, or both.
\item
  A correlation between $\alpha_{\rm ox}$ and $\dot{m}$ is predicted
  from several models of `slab'-type corona. We argue that the weaker
  dependence of $\alpha_{\rm ox}$ on $L_X / L_{Edd}$ implies that
  $L_{UV} / L_X$ increases with increasing $\dot{m}$. Considering that
  the efficiency of quasar feedback energy in driving an outflow may
  depend on the ratio of UV to X-ray luminosity, a correlation between
  $\alpha_{\rm ox}$ and both $M_{BH}$ and $\dot{m}$ has important
  consequences for models of black hole growth. In particular, if
  supermassive black holes become more X-ray quiet at higher
  $\dot{m}$, they will become more efficient at driving away their
  accreting gas, thus halting their growth.
\item
  Because of a possible nonlinear dependence of $\Gamma_X$ on $M_{BH},
  L_{UV} / L_{Edd}$, or $L_X / L_{Edd}$, we performed seperate
  regressions for the black hole mass estimates obtained from each
  emission line. We confirmed the significant dependence of $\Gamma_X$
  on $L_{UV} / L_{Edd}$ and $L_X / L_{Edd}$ seen in previous studies
  as inferred from the broad line mass estimates based on the H$\beta$
  line; however, we also find evidence that the $\Gamma_X$
  correlations change direction when including the C IV line. In
  particular, for the H$\beta$ sample, the X-ray continuum hardens
  with increasing $M_{BH}$, while for the C IV sample, the X-ray
  continuum softens with increasing $M_{BH}$. Similar but opposite
  trends are seen with respect to $L_{UV} / L_{Edd}$ and $L_X /
  L_{Edd}$, and we conclude that these relationships can be
  interpreted as resulting from a correlation between $\Gamma_X$ and
  $\dot{m}$. Results obtained from the Mg II line were too uncertain
  to interpret. We analyzed two test samples to argue that this
  non-monotonic behavior is not due to the different redshifts probed
  by the two samples, or to problems with the estimates of $M_{BH}$
  derived from the two lines; the different trends may be due to the
  difference in $M_{BH}$ probed by the two samples. A non-monotonic
  dependence of $\Gamma_X$ on $M_{BH}$ and/or $\dot{m}$ may imply a
  change in the disk/corona structure, although a non-monotonic
  dependence of $\Gamma_X$ on $\dot{m}$ and the viscosity is predicted
  by some models of `slab'-type coronal geometries.
\end{itemize}

\acknowledgements

We would like to thank the anonymous referee for a careful reading and
comments that contributed to the improvement of this paper. This
research was funded in part by NSF grant AST03-07384, NASA contract
NAS8-39073, and Chandra Award Number G05-6113X issued by the Chandra
X-ray Observatory Center. Funding for the SDSS and SDSS-II has been
provided by the Alfred P. Sloan Foundation, the Participating
Institutions, the National Science Foundation, the U.S. Department of
Energy, the National Aeronautics and Space Administration, the
Japanese Monbukagakusho, the Max Planck Society, and the Higher
Education Funding Council for England. The SDSS Web Site is
http://www.sdss.org/.

The SDSS is managed by the Astrophysical Research Consortium for the
Participating Institutions. The Participating Institutions are the
American Museum of Natural History, Astrophysical Institute Potsdam,
University of Basel, University of Cambridge, Case Western Reserve
University, University of Chicago, Drexel University, Fermilab, the
Institute for Advanced Study, the Japan Participation Group, Johns
Hopkins University, the Joint Institute for Nuclear Astrophysics, the
Kavli Institute for Particle Astrophysics and Cosmology, the Korean
Scientist Group, the Chinese Academy of Sciences (LAMOST), Los Alamos
National Laboratory, the Max-Planck-Institute for Astronomy (MPIA),
the Max-Planck-Institute for Astrophysics (MPA), New Mexico State
University, Ohio State University, University of Pittsburgh,
University of Portsmouth, Princeton University, the United States
Naval Observatory, and the University of Washington.

\begin{figure}
  \begin{center}
    \includegraphics[scale=0.33,angle=90]{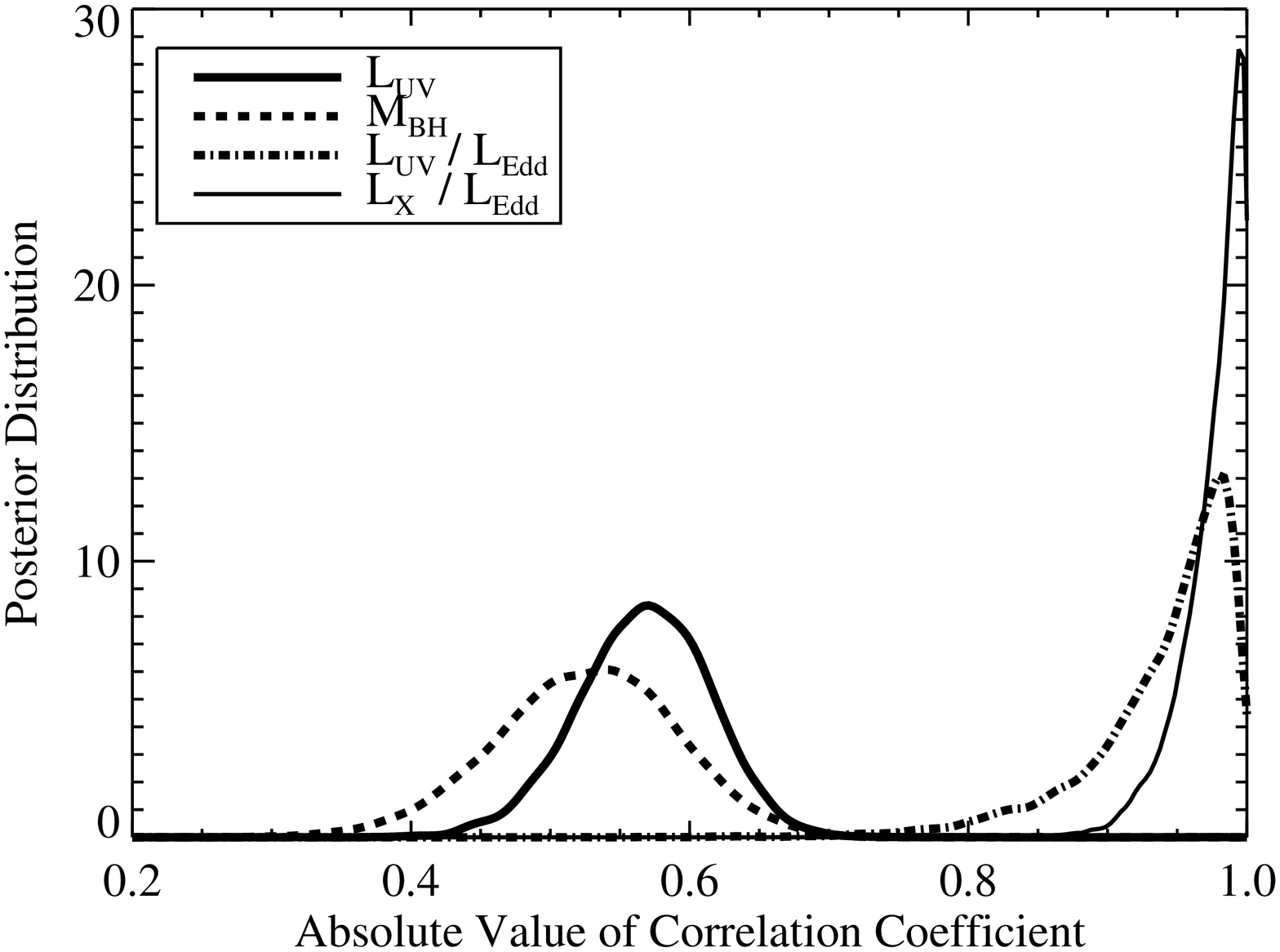}
    \includegraphics[scale=0.33,angle=90]{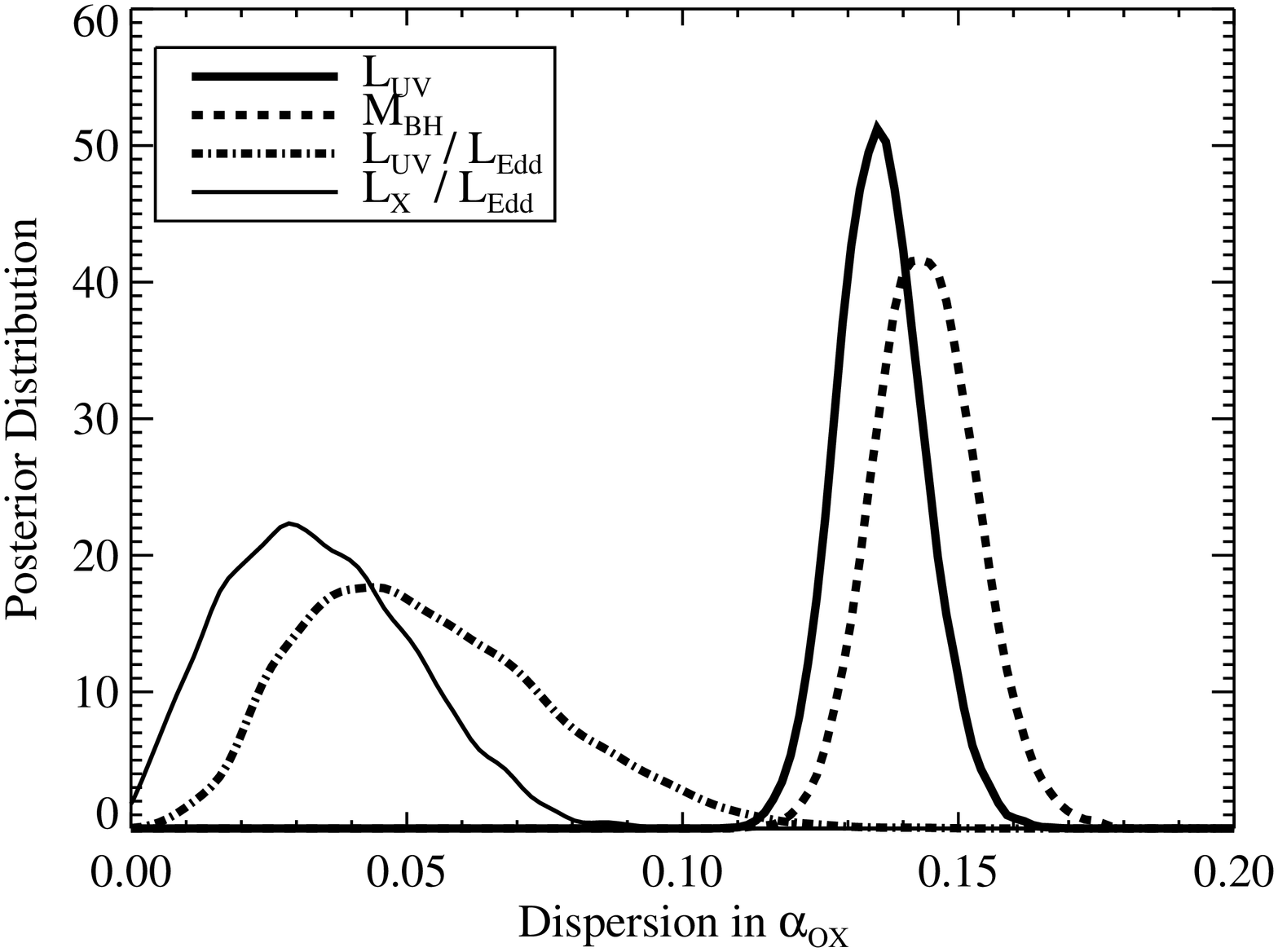}
    \caption{Posterior probability distributions for the absolute
    value of the correlation coefficients between $\alpha_{\rm ox}$
    and $\log L_{UV}, \log M_{BH}, \log L_{UV} / L_{Edd},$ and $\log
    L_X / L_{Edd}$ (left) and for the intrinsic dispersion in
    $\alpha_{\rm ox}$ at a given $L_{UV}, M_{BH}, L_{UV} / L_{Edd},$
    or $L_X / L_{Edd}$ (right). The thick solid line denotes the
    probability distribution for the two quantities with respect to
    $L_{UV}$, the dashed line for the two quantities with respect to
    $M_{BH}$, the dashed-dotted line for the two quantites with
    respect to $L_{UV} / L_{Edd}$, and the thin solid line for the two
    quantities with respect to $L_X / L_{Edd}$.
    \label{f-alfox_corr}}
  \end{center}
\end{figure}

\begin{deluxetable}{cccccc}
  \tabletypesize{\scriptsize}
  \tablecaption{Black Hole Parameters of the Sample\label{t-sample}}
  \tablewidth{0pt}
  \tablehead{
    \colhead{$\alpha$ (J2000)} 
    & \colhead{$\delta$ (J2000)} 
    & \colhead{$z$}
    & \colhead{$\log \hat{M}_{BL} / M_{\odot}$} 
    & \colhead{$\log L_{X} / \hat{L}_{Edd}$\tablenotemark{a}} 
    & \colhead{$\log L_{UV} / \hat{L}_{Edd}$\tablenotemark{b}} 
  }
  \startdata
  00 02 30.7 & +00 49 59.0 & 1.352 & $9.20 \pm 0.45$ & $-2.43 \pm 0.45$ & $-1.44 \pm 0.45$  \\
  00 06 54.1 & -00 15 33.4 & 1.725 & $9.08 \pm 0.29$ & $-2.58 \pm 0.30$ & $-1.07 \pm 0.29$  \\
  00 22 10.0 & +00 16 29.3 & 0.574 & $7.96 \pm 0.32$ & $-2.73 \pm 0.32$ & $-1.11 \pm 0.32$  \\
  00 27 52.4 & +00 26 15.7 & 0.205 & $7.35 \pm 0.45$ & $-2.42 \pm 0.46$ & $-1.26 \pm 0.49$  \\
  00 31 31.4 & +00 34 20.2 & 1.735 & $9.17 \pm 0.29$ & $-2.66 \pm 0.29$ & $-1.31 \pm 0.29$  \\
  00 50 06.3 & -00 53 19.0 & 4.331 & $9.61 \pm 0.41$ & $-2.89 \pm 0.42$ & $-1.19 \pm 0.41$  \\
  00 57 01.1 & +14 50 03.0 & 0.623 & $8.66 \pm 0.32$ & $-3.79 \pm 0.32$ & $-1.66 \pm 0.32$  \\
  00 59 22.8 & +00 03 01.0 & 4.178 & $9.21 \pm 0.38$ & $-2.28 \pm 0.41$ & $-0.74 \pm 0.38$  \\
  01 06 19.2 & +00 48 22.0 & 4.437 & $9.18 \pm 0.38$ & $-1.96 \pm 0.39$ & $-0.46 \pm 0.38$  \\
  01 13 05.7 & +15 31 46.5 & 0.576 & $9.39 \pm 0.32$ & $-3.47 \pm 0.32$ & $-2.46 \pm 0.32$  \\
  01 13 09.1 & +15 35 53.6 & 1.806 & $9.19 \pm 0.29$ & $-2.32 \pm 0.29$ & $-1.13 \pm 0.29$  \\
  01 15 37.7 & +00 20 28.7 & 1.275 & $9.25 \pm 0.45$ & $-2.75 \pm 0.45$ & $-1.81 \pm 0.45$  \\
  01 26 02.2 & -00 19 24.1 & 1.765 & $8.96 \pm 0.29$ & $-2.58 \pm 0.29$ & $-1.01 \pm 0.29$  \\
  \enddata
  
  \tablecomments{The complete version of this table is in the
    electronic edition of the Journal. The printed edition contains
    only a sample.}
  \tablenotetext{a}{Logarithm of the ratio of $\nu L_{\nu}$ [2 keV] to
    $\hat{L}_{Edd}$, where $\hat{L}_{Edd}$ is calculated from the broad
    emission line estimate of $M_{BH}, \hat{M}_{BL}$.}
  \tablenotetext{b}{Logarithm of the ratio of $\nu L_{\nu}$ [2500\AA] to
    $\hat{L}_{Edd}$, where $\hat{L}_{Edd}$ is calculated from the broad
    emission line estimate of $M_{BH}, \hat{M}_{BL}$.}
  
\end{deluxetable}


\begin{deluxetable}{lccccccccc}
  \tabletypesize{\scriptsize}
  \tablecaption{Sources with H$\beta$ and C IV Used for Testing the $\Gamma_X$--$M_{BH}$ 
    Correlations \label{t-test}}
  \tablewidth{0pt}
  \tablehead{
    \colhead{Quasar Name} & \colhead{$\alpha$} & \colhead{$\delta$} & \colhead{$z$} & \colhead{Line} & 
    \colhead{$\log \hat{M}_{BL}$} & \colhead{$\hat{L}_{bol} / \hat{L}_{Edd}$} & Opt. Ref.\tablenotemark{a} & 
    $\Gamma_X$ & X-ray Ref.\tablenotemark{b} \\
    \colhead{} & \colhead{(J2000)} & \colhead{(J2000)} & \colhead{} & \colhead{} & \colhead{$M_{\odot}$} &
    \colhead{} & \colhead{} & \colhead{} & \colhead{}
  }
  \startdata
  PG 0026+129      & 00 29 13.7 & +13 16 03.8 & 0.142 & C IV     & 7.789 & 1.768 & 1 & 1.96\tablenotemark{c} & 5 \\
  Fairall 9        & 01 23 45.7 & -58 48 21.8 & 0.046 & C IV     & 7.760 & 0.265 & 1 & $1.83 \pm 0.06$ & 6 \\
  PG 1202+281      & 12 04 42.2 & +27 54 12.0 & 0.165 & C IV     & 7.855 & 0.359 & 1 & $1.76 \pm 0.07$ & 7 \\
  PG 1211+143      & 12 14 17.7 & +14 03 12.3 & 0.080 & C IV     & 7.559 & 1.380 & 1 & $2.06 \pm 0.05$ & 8 \\
  PG 1247+267      & 12 50 05.7 & +26 31 07.7 & 2.038 & H$\beta$ & 10.41 & 0.379 & 2 & $2.23 \pm 0.10$ & 9 \\
  Q1346-036        & 13 48 44.1 & -03 53 25.0 & 2.370 & H$\beta$ & 9.946 & 0.609 & 3 & $2.02 \pm 0.17$ & 3 \\
  MRK 478          & 14 42 07.5 & +35 26 22.9 & 0.077 & C IV     & 7.890 & 0.498 & 1 & $2.41 \pm 0.07$ & 7 \\ 
  PG 1630+377      & 16 32 01.1 & +37 37 50.0 & 1.476 & H$\beta$ & 9.762 & 0.569 & 4 & $2.20 \pm 0.30$ & 10 \\
  PG 1634+706      & 16 34 28.9 & +70 31 33.0 & 1.334 & H$\beta$ & 10.27 & 0.734 & 4 & $2.19 \pm 0.05$ & 10 \\
  HE 2217-2818     & 22 20 06.8 & -28 03 23.9 & 2.414 & H$\beta$ & 10.12 & 0.807 & 3 & $1.97 \pm 0.06$ & 3
  \enddata

  \tablenotetext{a}{Reference for the rest frame optical/UV data.}
  \tablenotetext{b}{Reference for the X-ray data.}
  \tablenotetext{c}{\citet{oneill05} do not report an error on $\Gamma_X$.}

  \tablerefs{
    (1) \citet{kelly07b} (2) \citet{mcint99} (3) \citet{shemmer06} (4) \citet{nishi97}
    (5) \citet{oneill05} (6) \citet{nandra97} (7) \citet{brock06} (8) \citet{reeves00}
    (9) \citet{page04} (10) \citet{picon05}
  }

\end{deluxetable}


\begin{deluxetable}{ccccccc}
  \tabletypesize{\scriptsize}
  \tablecaption{Test Sources from COSMOS \label{t-cosmos}}
  \tablewidth{0pt}
  \tablehead{
    \colhead{$\alpha$} & \colhead{$\delta$} & \colhead{$z$} & \colhead{Line} & 
    \colhead{$\log \hat{M}_{BL}$} & \colhead{$\hat{L}_{bol} / \hat{L}_{Edd}$} & $\Gamma_X$ \\
    \colhead{(J2000)} & \colhead{(J2000)} & \colhead{} & \colhead{} & \colhead{$M_{\odot}$} &
    \colhead{} & \colhead{}
  }
  \startdata
  09 58 48.8 & +02 34 42.3 & 1.551 & C IV  & 8.276 & 0.131 & $2.01 \pm 0.11$ \\
  09 59 02.6 & +02 25 11.8 & 1.105 & Mg II & 7.612 & 0.025 & $2.17 \pm 0.28$ \\
  09 59 49.4 & +02 01 41.1 & 1.758 & C IV  & 8.108 & 0.719 & $2.51 \pm 0.16$ \\
  10 00 50.0 & +02 05 00.0 & 1.235 & Mg II & 7.692 & 0.501 & $2.50 \pm 0.13$ \\
  10 00 51.6 & +02 12 15.8 & 1.829 & Mg II & 7.807 & 0.131 & $2.14 \pm 0.17$ \\
  10 00 58.9 & +01 53 59.5 & 1.559 & C IV  & 8.346 & 0.172 & $2.04 \pm 0.17$ \\
  10 02 19.6 & +01 55 36.9 & 1.509 & C IV  & 8.333 & 0.177 & $2.19 \pm 0.23$ \\
  10 02 34.4 & +01 50 11.5 & 1.506 & C IV  & 7.991 & 0.941 & $2.25 \pm 0.12$ \\
  10 02 43.9 & +02 05 02.0 & 1.234 & Mg II & 7.817 & 0.303 & $1.97 \pm 0.29$ \\
  \enddata

\end{deluxetable}


\begin{thebibliography}{}
\begin{scriptsize}

\bibitem[Abramowicz et al.(1988)]{abram88} Abramowicz, M.~A., Czerny,
  B., Lasota, J.~P., \& Szuszkiewicz, E.\ 1988, \apj, 332, 646
\bibitem[Akritas \& Bershady(1996)]{bces} Akritas, M.~G., \& 
  Bershady, M.~A.\ 1996, \apj, 470, 706 
\bibitem[Anderson et al.(2001)]{and01} Anderson, S.~F., et 
  al.\ 2001, \aj, 122, 503 
\bibitem[Avni \& Tananbaum(1982)]{avni82} Avni, Y., \& Tananbaum, H.\
  1982, \apjl, 262, L17
\bibitem[Ballantyne et al.(2001)]{ball01} Ballantyne, D.~R., Ross,
  R.~R., \& Fabian, A.~C.\ 2001, \mnras, 327, 10
\bibitem[Bechtold et al.(2003)]{bech03} Bechtold, J., et
al.\ 2003, \apj, 588, 119
\bibitem[Begelman \& Nath(2005)]{begel05} Begelman, M.~C., \& Nath,
B.~B.\ 2005, \mnras, 361, 1387
\bibitem[Bentz et al.(2006)]{bentz06} Bentz, M.~C., Peterson, B.~M.,
  Pogge, R.~W., Vestergaard, M., \& Onken, C.~A.\ 2006, \apj, 644, 133
\bibitem[Bian(2005)]{bian05} Bian, W.-H.\ 2005, Chinese Journal of
  Astronomy and Astrophysics Supplement, 5, 289
\bibitem[Bianchi et al.(2007)]{bianchi07} Bianchi, S., Guainazzi, M.,
  Matt, G., \& Fonseca Bonilla, N.\ 2007, \aap, 467, L19
\bibitem[Bisnovatyi-Kogan \& Blinnikov(1977)]{bis77} Bisnovatyi-Kogan,
  G.~S., \& Blinnikov, S.~I.\ 1977, \aap, 59, 111
\bibitem[Boller et al.(1996)]{boller96} Boller, T., Brandt, W.~N., \&
  Fink, H.\ 1996, \aap, 305, 53
\bibitem[Bonning et al.(2007)]{bonning07} Bonning, E.~W., Cheng, L.,
  Shields, G.~A., Salviander, S., \& Gebhardt, K.\ 2007, \apj, 659,
  211
\bibitem[Brandt et al.(1997)]{brandt97} Brandt, W.~N., Mathur, S., \&
  Elvis, M.\ 1997, \mnras, 285, L25
\bibitem[Brandt \& Boller(1998)]{brandt98} Brandt, N., \& Boller, T.\
  1998, Astronomische Nachrichten, 319, 163
\bibitem[Brocksopp et al.(2006)]{brock06} Brocksopp, C., 
  Starling, R.~L.~C., Schady, P., Mason, K.~O., Romero-Colmenero, E., \& 
  Puchnarewicz, E.~M.\ 2006, \mnras, 366, 953 
\bibitem[Bromley et al.(2004)]{bromley04} Bromley, J.~M., Somerville,
  R.~S., \& Fabian, A.~C.\ 2004, \mnras, 350, 456
\bibitem[Brunner et al.(1997)]{brunner97} Brunner, H., Mueller, C.,
  Friedrich, P., Doerrer, T., Staubert, R., \& Riffert, H.\ 1997, \aap,
  326, 885
\bibitem[Budav{\'a}ri et al.(2005)]{bud05} Budav{\'a}ri, T., 
  et al.\ 2005, \apjl, 619, L31 
\bibitem[Cardelli et al.(1989)]{ccm89} Cardelli, J.~A., 
  Clayton, G.~C., \& Mathis, J.~S.\ 1989, \apj, 345, 245
\bibitem[Collin et al.(2006)]{collin06} Collin, S., Kawaguchi, T.,
  Peterson, B.~M., \& Vestergaard, M.\ 2006, \aap, 456, 75
\bibitem[Congdon(2006)]{congdon06} Congdon, P.\ 2006, Bayesian
  Statistical Modelling (2nd. Ed.; West Sussex:John Wiley \& Sons
  Ltd.)
\bibitem[Constantin et al.(2002)]{const02} Constantin, A., Shields,
  J.~C., Hamann, F., Foltz, C.~B., \& Chaffee, F.~H.\ 2002, \apj, 565,
  50
\bibitem[Czerny et al.(2003)]{czerny03} Czerny, B., Niko{\l}ajuk, M.,
  R{\'o}{\.z}a{\'n}ska, A., Dumont, A.-M., Loska, Z., \& Zycki, P.~T.\
  2003, \aap, 412, 317
\bibitem[Czerny et al.(1997)]{czerny97} Czerny, B., Witt, H.~J., 
\&
  Zycki, P.\ 1997, The Transparent Universe, 382, 397
\bibitem[Dai et al.(2004)]{dai04} Dai, X., Chartas, G., Eracleous, M.,
  \& Garmire, G.~P.\ 2004, \apj, 605, 45
\bibitem[Davis et al.(2007)]{davis07} Davis, S.~W., Woo, J.-H., \&
  Blaes, O.~M.\ 2007, \apj, 668, 682
\bibitem[Dempster, Laird, \& Rubin(1977)]{em} Dempster, A., Laird, N.,
  \& Rubin, D.\ 1977, J.~R. Statist. Soc. B., 39, 1
\bibitem[De Villiers et al.(2003)]{devill03} De Villiers, J.-P.,
  Hawley, J.~F., \& Krolik, J.~H.\ 2003, \apj, 599, 1238
\bibitem[Di Matteo et al.(2005)]{dimatt05} Di Matteo, T., Springel,
  V., \& Hernquist, L.\ 2005, \nat, 433, 604
\bibitem[Elvis et al.(2002)]{elvis02} Elvis, M., Risaliti, G., \&
  Zamorani, G.\ 2002, \apjl, 565, L75
\bibitem[Elvis et al.(1994)]{elvis94} Elvis, M., et al.\ 1994, \apjs,
  95, 1
\bibitem[Esin et al.(1997)]{esin97} Esin, A.~A., McClintock, J.~E., \&
  Narayan, R.\ 1997, \apj, 489, 865
\bibitem[Fabian(1999)]{fabian99} Fabian, A.~C.\ 1999, \mnras, 308, L39
\bibitem[Fabian et al.(2002)]{fabian02} Fabian, A.~C., et al.\ 2002,
  \mnras, 335, L1
\bibitem[Fox(1997)]{fox97} Fox, J.\ 1997, Applied Regression
  Analysis, Linear Models, and Related Methods (Thousand Oaks:Sage
  Publications, Inc.)
\bibitem[Fuller(1987)]{fuller87} Fuller, W.~A.\ 1987, Measurement
  Error Models (New York:John Wiley \& Sons)
\bibitem[Galeev et al.(1979)]{gal79} Galeev, A.~A., Rosner, R., \&
  Vaiana, G.~S.\ 1979, \apj, 229, 318
\bibitem[Gallagher et al.(2002)]{gall02} Gallagher, S.~C., Brandt,
  W.~N., Chartas, G., \& Garmire, G.~P.\ 2002, \apj, 567, 37
\bibitem[Gallagher et al.(2005)]{gall05} Gallagher, S.~C., 
  Richards, G.~T., Hall, P.~B., Brandt, W.~N., Schneider, D.~P., \& Vanden 
  Berk, D.~E.\ 2005, \aj, 129, 567 
\bibitem[Gallagher et al.(2006)]{gall06} Gallagher, S.~C., Brandt,
  W.~N., Chartas, G., Priddey, R., Garmire, G.~P., \& Sambruna, R.~M.\
  2006, \apj, 644, 709
\bibitem[Gebhardt et al.(2000)]{gebh00} Gebhardt, K., et al.\ 
  2000, \apjl, 539, L13 
\bibitem[Gierli{\'n}ski \& Done(2004)]{gier04} Gierli{\'n}ski, M., \&
  Done, C.\ 2004, \mnras, 349, L7
\bibitem[Gilli et al.(2007)]{gilli07} Gilli, R., Comastri, A., \&
  Hasinger, G.\ 2007, \aap, 463, 79
\bibitem[Green et al.(2001)]{green01} Green, P.~J., Aldcroft, T.~L.,
  Mathur, S., Wilkes, B.~J., \& Elvis, M.\ 2001, \apj, 558, 109
\bibitem[Grupe et al.(2006)]{grupe06} Grupe, D., Mathur, S., 
  Wilkes, B., \& Osmer, P.\ 2006, \aj, 131, 55 
\bibitem[Grupe et al.(2001)]{grupe01} Grupe, D., Thomas, H.-C., 
  \& Beuermann, K.\ 2001, \aap, 367, 470 
\bibitem[Haardt \& Maraschi(1991)]{haardt91} Haardt, F., \& Maraschi,
  L.\ 1991, \apjl, 380, L51
\bibitem[Haardt \& Maraschi(1993)]{haardt93} Haardt, F., \&
  Maraschi, L.\ 1993, \apj, 413, 507
\bibitem[Hastie, Tibshirani, \& Friedman(2001)]{hastie01} Hastie, T.,
  Tibshirani, R., \& Friedman, J.\ 2001, The Elements of Statistical
  Learning (New York:Springer-Verlag)
\bibitem[Hatziminaoglou et al.(2003)]{hatz03} Hatziminaoglou, E.,
  Mathez, G., Solanes, J.-M., Manrique, A., \& Salvador-Sol{\'e}, E.\
  2003, \mnras, 343, 692
\bibitem[Hopkins et al.(2005)]{hopkins05} Hopkins, P.~F.,
  Hernquist, L., Cox, T.~J., Di Matteo, T., Martini, P., Robertson, B.,
  \& Springel, V.\ 2005, \apj, 630, 705
\bibitem[Hopkins et al.(2006)]{hopkins06} Hopkins, P.~F.,
  Hernquist, L., Cox, T.~J., Di Matteo, T., Robertson, B., \& Springel,
  V.\ 2006, \apjs, 163, 1
\bibitem[Hopkins et al.(2007)]{hopkins07} Hopkins, P.~F.,
  Richards, G.~T., \& Hernquist, L.\ 2007, \apj, 654, 731
\bibitem[Janiuk \& Czerny(2000)]{janiuk00} Janiuk, A., \& Czerny, B.\
  2000, New Astronomy, 5, 7
\bibitem[Janiuk et al.(2001)]{janiuk01} Janiuk, A., Czerny, B., \&
  Madejski, G.~M.\ 2001, \apj, 557, 408
\bibitem[Just et al.(2007)]{just07} Just, D.~W., Brandt, 
  W.~N., Shemmer, O., Steffen, A.~T., Schneider, D.~P., Chartas, G., \& 
  Garmire, G.~P.\ 2007, \apj, 665, 1004 
\bibitem[Kaspi et al.(2005)]{kaspi05} Kaspi, S., Maoz, D., 
  Netzer, H., Peterson, B.~M., Vestergaard, M., \& Jannuzi, B.~T.\ 2005, 
  \apj, 629, 61 
\bibitem[Kauffmann \& Haehnelt(2000)]{kauff00} Kauffmann, G., \&
  Haehnelt, M.\ 2000, \mnras, 311, 576
\bibitem[Kelly(2007)]{kelly07a} Kelly, B.~C.\ 2007, \apj, 665, 
  1489 
\bibitem[Kelly \& Bechtold(2007)]{kelly07b} Kelly, B.~C., \& 
  Bechtold, J.\ 2007, \apjs, 168, 1 
\bibitem[Kelly et al.(2007)]{kelly07c} Kelly, B.~C., Bechtold, 
  J., Siemiginowska, A., Aldcroft, T., \& Sobolewska, M.\ 2007, \apj, 657, 
  116 (K07)
\bibitem[Kollmeier et al.(2006)]{koll06} Kollmeier, J.~A., et al.\
  2006, \apj, 648, 128
\bibitem[Krolik(2001)]{krolik01} Krolik, J.~H.\ 2001, \apj, 551, 72
\bibitem[Krolik et al.(2005)]{krolik05} Krolik, J.~H., Hawley, J.~F.,
  \& Hirose, S.\ 2005, \apj, 622, 1008
\bibitem[Laor et al.(1997)]{laor97} Laor, A., Fiore, F., Elvis, M.,
  Wilkes, B.~J., \& McDowell, J.~C.\ 1997, \apj, 477, 93
\bibitem[Lee et al.(2002)]{lee02} Lee, J.~C., Iwasawa, K., Houck,
  J.~C., Fabian, A.~C., Marshall, H.~L., \& Canizares, C.~R.\ 2002,
  \apjl, 570, L47
\bibitem[Liang \& Price(1977)]{liang77} Liang, E.~P.~T., \& Price,
  R.~H.\ 1977, \apj, 218, 247
\bibitem[Liu et al.(2003)]{liu03} Liu, B.~F., Mineshige, S., \&
  Ohsuga, K.\ 2003, \apj, 587, 571
\bibitem[Lu \& Yu(1999)]{lu99} Lu, Y., \& Yu, Q.\ 1999, \apjl, 526, L5
\bibitem[Mainieri(2007)]{mainieri07} Mainieri, V.\ 2007, in press at
  ApJS (astro-ph/0612361)
\bibitem[Malkan(1991)]{malkan91} Malkan, M.\ 1991, IAU Colloq.~129:
  The 6th Institute d'Astrophysique de Paris (IAP) Meeting:
  Structure and Emission Properties of Accretion Disks, 165
\bibitem[Markowitz et al.(2003)]{mark03} Markowitz, A., Edelson, R.,
  \& Vaughan, S.\ 2003, \apj, 598, 935
\bibitem[Mathur(2000)]{mathur00} Mathur, S.\ 2000, \mnras, 314, 
  L17 
\bibitem[McIntosh et al.(1999)]{mcint99} McIntosh, D.~H., Rieke, M.~J., 
  Rix, H.-W., Foltz, C.~B., \& Weymann, R.~J.\ 1999, \apj, 514, 40
\bibitem[McLure \& Dunlop(2004)]{mclure04} McLure, R.~J., \& Dunlop,
  J.~S.\ 2004, \mnras, 352, 1390
\bibitem[McLure \& Jarvis(2002)]{mclure02} McLure, R.~J., \& 
  Jarvis, M.~J.\ 2002, \mnras, 337, 109 
\bibitem[Merloni \& Fabian(2002)]{merloni02} Merloni, A., \&
  Fabian, A.~C.\ 2002, \mnras, 332, 165
\bibitem[Merritt \& Ferrarese(2001)]{merr01} Merritt, D., \& 
  Ferrarese, L.\ 2001, \apj, 547, 140 
\bibitem[Mineo et al.(2000)]{mineo00} Mineo, T., et al.\
  2000, \aap, 359, 471
\bibitem[Murray et al.(1995)]{murray95} Murray, N., Chiang,
  J., Grossman, S.~A., \& Voit, G.~M.\ 1995, \apj, 451, 498
\bibitem[Nandra et al.(1997)]{nandra97} Nandra, K., George, I.~M.,
  Mushotzky, R.~F., Turner, T.~J., \& Yaqoob, T.\ 1997, \apj, 477, 602
\bibitem[Nayakshin(2000)]{nayak00} Nayakshin, S.\ 2000, \apj, 534, 718
\bibitem[Nishihara et al.(1997)]{nishi97} Nishihara, E., Yamashita,
  T., Yoshida, M., Watanabe, E., Okumura, S.-I., Mori, A., \& Iye, M.\
  1997, \apjl, 488, L27
\bibitem[Nowak et al.(2002)]{nowak02} Nowak, M.~A., Wilms, J., \&
  Dove, J.~B.\ 2002, \mnras, 332, 856
\bibitem[O'Neill et al.(2005)]{oneill05} O'Neill, P.~M., Nandra, 
  K., Papadakis, I.~E., \& Turner, T.~J.\ 2005, \mnras, 358, 1405 
\bibitem[Onken et al.(2004)]{onken04} Onken, C.~A., Ferrarese, 
  L., Merritt, D., Peterson, B.~M., Pogge, R.~W., Vestergaard, M., \& Wandel, 
  A.\ 2004, \apj, 615, 645 
\bibitem[Page et al.(2004)]{page04} Page, K.~L., Reeves, 
  J.~N., O'Brien, P.~T., Turner, M.~J.~L., \& Worrall, D.~M.\ 2004, \mnras, 
  353, 133 
\bibitem[Page et al.(2003)]{page03} Page, K.~L., Turner, M.~J.~L.,
  Reeves, J.~N., O'Brien, P.~T., \& Sembay, S.\ 2003, \mnras, 338,
  1004
\bibitem[P{\' e}roux et al.(2001)]{peroux01} P{\' e}roux, C., 
  Storrie-Lombardi, L.~J., McMahon, R.~G., Irwin, M., \& Hook, I.~M.\ 2001, 
  \aj, 121, 1799 
\bibitem[Pessah et al.(2006)]{pessah06} Pessah, M.~E.,
  Chan, C.-K., \& Psaltis, D.\ 2006, \mnras, 372, 183
\bibitem[Pessah et al.(2007)]{pessah07} Pessah, M.~E., Chan, C.-k., \&
  Psaltis, D.\ 2007, in press at \apj (arXiv:0705.0352)
\bibitem[Peterson et al.(2004)]{peter04} Peterson, B.~M., et al.\ 2004, 
  \apj, 613, 682 
\bibitem[Piconcelli et al.(2005)]{picon05} Piconcelli, E., 
  Jimenez-Bail{\'o}n, E., Guainazzi, M., Schartel, N., 
  Rodr{\'{\i}}guez-Pascual, P.~M., \& Santos-Lle{\'o}, M.\ 2005, \aap, 432, 
  15 
\bibitem[Porquet et al.(2004)]{porquet04} Porquet, D., Reeves, J.~N.,
  O'Brien, P., \& Brinkmann, W.\ 2004, \aap, 422, 85
\bibitem[Poutanen et al.(1997)]{pout97} Poutanen, J., Krolik, J.~H.,
  \& Ryde, F.\ 1997, \mnras, 292, L21
\bibitem[Proga(2005)]{proga05} Proga, D.\ 2005, \apjl, 630, L9
\bibitem[Proga(2007)]{proga07} Proga, D.\ 2007, \apj, 661, 693
\bibitem[Proga \& Kallman(2004)]{proga04} Proga, D., \&
  Kallman, T.~R.\ 2004, \apj, 616, 688
\bibitem[Reeves \& Turner(2000)]{reeves00} Reeves, J.~N., \& 
  Turner, M.~J.~L.\ 2000, \mnras, 316, 234 
\bibitem[Reeves et al.(1997)]{reeves97} Reeves, J.~N., Turner,
  M.~J.~L., Ohashi, T., \& Kii, T.\ 1997, \mnras, 292, 468
\bibitem[Reichard et al.(2003)]{reich03} Reichard, T.~A., et al.\
  2003, \aj, 126, 2594
\bibitem[Richards et al.(2006)]{rich06} Richards, G.~T., et al.\ 2006,
  \aj, 131, 2766
\bibitem[Risaliti \& Elvis(2005)]{ris05} Risaliti, G., \& Elvis, M.\
  2005, \apjl, 629, L17
\bibitem[Schlegel et al.(1998)]{schlegel} Schlegel, D.~J., Finkbeiner,
  D.~P., \& Davis, M.\ 1998, \apj, 500, 525
\bibitem[Schneider et al.(2005)]{dr3qsos} Schneider, D.~P., et al.\
  2005, \aj, 130, 376
\bibitem[Schwartz(1979)]{bic} Schwartz, G.\ 1979, Ann.~Statist., 6, 461
\bibitem[Scoville(2007)]{cosmos} Scoville, N.~Z., et al.\ 2007, in
  press at \apjs (astro-ph/0612305)
\bibitem[Shakura \& Syunyaev(1973)]{shakura73} Shakura, N.~I., \&
  Syunyaev, R.~A.\ 1973, \aap, 24, 337
\bibitem[Shapiro et al.(1976)]{shap76} Shapiro, S.~L., Lightman,
  A.~P., \& Eardley, D.~M.\ 1976, \apj, 204, 187
\bibitem[Shemmer et al.(2006)]{shemmer06} Shemmer, O., Brandt, 
  W.~N., Netzer, H., Maiolino, R., \& Kaspi, S.\ 2006, \apjl, 646, L29 
\bibitem[Silk \& Rees(1998)]{silk98} Silk, J., \& Rees, M.~J.\ 1998,
  \aap, 331, L1
\bibitem[Sobolewska et al.(2004a)]{sob04a} Sobolewska, M.~A.,
  Siemiginowska, A., \& \.{Z}ycki, P.~T.\ 2004a, \apj, 608, 80
\bibitem[Sobolewska et al.(2004b)]{sob04b} Sobolewska, M.~A.,
  Siemiginowska, A., \& \.{Z}ycki, P.~T.\ 2004b, \apj, 617, 102
\bibitem[Spergel et al.(2003)]{wmap} Spergel, D.~N., et al.\ 
  2003, \apjs, 148, 175 
\bibitem[Springel et al.(2005)]{spring05} Springel, V., Di Matteo, T.,
  \& Hernquist, L.\ 2005, \apjl, 620, L79
\bibitem[Stapleton \& Young(1984)]{stap84} Stapleton, D.~C. \& Young,
  D.~J.\ 1984, Econometrica, 52, 737
\bibitem[Steffen et al.(2006)]{steffen06} Steffen, A.~T., 
  Strateva, I., Brandt, W.~N., Alexander, D.~M., Koekemoer, A.~M., Lehmer, 
  B.~D., Schneider, D.~P., \& Vignali, C.\ 2006, \aj, 131, 2826
\bibitem[Storrie-Lombardi et al.(1996)]{storr96} Storrie-Lombardi,
  L.~J., McMahon, R.~G., Irwin, M.~J., \& Hazard, C.\ 1996, \apj, 468,
  121
\bibitem[Strateva et al.(2005)]{strat05} Strateva, I.~V., Brandt,
  W.~N., Schneider, D.~P., Vanden Berk, D.~G., \& Vignali, C.\ 2005,
  \aj, 130, 387 (S05)
\bibitem[Tananbaum et al.(1979)]{tan79} Tananbaum, H., et 
  al.\ 1979, \apjl, 234, L9 
\bibitem[Turner(2004)]{turner04} Turner, N.~J.\ 2004, \apjl, 605, L45
\bibitem[Tremaine et al.(2002)]{trem02} Tremaine, S., et al.\ 
  2002, \apj, 574, 740 
\bibitem[Trump et al.(2006)]{trump06} Trump, J.~R., et
  al.\ 2006, \apjs, 165, 1
\bibitem[Trump et al.(2007)]{trump07} Trump, J., Impey, C., et al.\
  2007, in press at \apjs (astro-ph/0606016)
\bibitem[Trump et al.(2008)]{trump08} Trump, J., Impey, C., et al.\
  2008, in preparation
\bibitem[Uttley et al.(2002)]{uttley02} Uttley, P., McHardy, I.~M., \&
  Papadakis, I.~E.\ 2002, \mnras, 332, 231
\bibitem[Vasudevan \& Fabian(2007)]{vasud07} Vasudevan, R.~V., \&
  Fabian, A.~C.\ 2007, in press at \mnras, (arXiv:0708.4308)
\bibitem[V{\' e}ron-Cetty et al.(2004)]{optfe} V{\' 
  e}ron-Cetty, M.-P., Joly, M., \& V{\' e}ron, P.\ 2004, \aap, 417, 515 
\bibitem[Vestergaard(2002)]{vest02} Vestergaard, M.\ 2002, \apj, 571,
  733
\bibitem[Vestergaard(2004)]{vest04} Vestergaard, M.\ 2004, \apj, 601,
  676
\bibitem[Vestergaard et al.(2008)]{vest08} Vestergaard, M., et al.\ 2007, in
  preparation
\bibitem[Vestergaard \& Peterson(2006)]{vest06} Vestergaard, 
  M., \& Peterson, B.~M.\ 2006, \apj, 641, 689 
\bibitem[Vestergaard \& Wilkes(2001)]{uvfe} Vestergaard, M., \&
  Wilkes, B.~J.\ 2001, \apjs, 134, 1
\bibitem[Vignali et al.(2001)]{vig01} Vignali, C., Brandt, W.~N., Fan,
  X., Gunn, J.~E., Kaspi, S., Schneider, D.~P., \& Strauss, M.~A.\
  2001, \aj, 122, 2143
\bibitem[Vignali et al.(2003a)]{vig03a} Vignali, C., Brandt, W.~N.,
  Schneider, D.~P., Garmire, G.~P., \& Kaspi, S.\ 2003, \aj, 125, 418
\bibitem[Vignali et al.(2003b)]{vig03b} Vignali, C., Brandt, W.~N., \&
  Schneider, D.~P.\ 2003, \aj, 125, 433
\bibitem[Vignali et al.(2005)]{vig05} Vignali, C., Brandt, W.~N.,
  Schneider, D.~P., \& Kaspi, S.\ 2005, \aj, 129, 2519
\bibitem[Vignali et al.(1999)]{vig99} Vignali, C., Comastri, A.,
  Cappi, M., Palumbo, G.~G.~C., Matsuoka, M., \& Kubo, H.\ 1999, \apj,
  516, 582
\bibitem[Wandel(2000)]{wandel00} Wandel, A.\ 2000, New Astronomy
  Review, 44, 427
\bibitem[Wandel et al.(1999)]{wand99} Wandel, A., Peterson, 
  B.~M., \& Malkan, M.~A.\ 1999, \apj, 526, 579
\bibitem[Wang et al.(2004)]{wang04} Wang, J.-M., Watarai, K.-Y., \&
  Mineshige, S.\ 2004, \apjl, 607, L107
\bibitem[Wilkes \& Elvis(1987)]{wilkes87} Wilkes, B.~J., \& 
  Elvis, M.\ 1987, \apj, 323, 243 
\bibitem[Wilkes et al.(1994)]{wilkes94} Wilkes, B.~J., Tananbaum, H.,
  Worrall, D.~M., Avni, Y., Oey, M.~S., \& Flanagan, J.\ 1994, \apjs,
  92, 53
\bibitem[Woo \& Urry(2002)]{woo02} Woo, J., \& Urry, C.~M.\ 2002,
  \apj, 579, 530
\bibitem[Worrall et al.(1987)]{worrall87} Worrall, D.~M., 
  Tananbaum, H., Giommi, P., \& Zamorani, G.\ 1987, \apj, 313, 596 
\bibitem[Wyithe \& Loeb(2003)]{wyithe03} Wyithe, J.~S.~B., \& Loeb,
  A.\ 2003, \apj, 595, 614
\bibitem[Yuan et al.(1998)]{yuan98} Yuan, W., Siebert, J., \& 
  Brinkmann, W.\ 1998b, \aap, 334, 498 
\bibitem[Zamorani et al.(1981)]{zam81} Zamorani, G., et al.\ 
  1981, \apj, 245, 357 
\bibitem[Zdziarski et al.(1999)]{zdz99} Zdziarski, A.~A., 
  Lubinski, P., \& Smith, D.~A.\ 1999, \mnras, 303, L11 

\end{scriptsize}

\end{thebibliography}
\end{document}